\documentclass[a4paper,11pt]{article}
\pdfoutput=1

\usepackage{jcappub} 

\usepackage[T1]{fontenc} 
\usepackage{gensymb}
\usepackage{graphicx}
\usepackage{caption}
\usepackage{subcaption}

\title{A Giant Ring on the sky}

\author[a]{A. M. Lopez,}
\author[a]{R. G. Clowes}

\affiliation[a]{Jeremiah Horrocks Institute,\\University of Lancashire,
Preston, PR1 2HE, United Kingdom}

\emailAdd{amlopez2@lancashire.ac.uk}
\emailAdd{rgclowes@lancashire.ac.uk}

\abstract{
We present the discovery of `A Giant Ring on the Sky' (GR); a ring-like, ultra-large-scale structure (uLSS) at $z \sim 0.8$, located in the same field that contains the previously-documented Giant Arc (GA) and Big Ring (BR) uLSSs.
The GR was \emph{predicted} from the presence of a Northern Arc (NA) filament (noted in previous work), which looked like it could, with more or enhanced data, connect with the GA to form a giant ring that encompasses the BR. 
There is now substantial evidence to support the reality of a GR.

There appear to be two overlapping versions of the GR which differ by only the left-hand-side (LHS, in the presented figures --- roughly western side) trajectory; this branching in the LHS of the GR was identified with the FilFinder algorithm and appears to correspond to both the GR prediction (the extended, elliptical, GR from the GA+NA ellipse), and the visually-identified ellipse (the visually-impressive, almost contiguous, roughly circular, GR which is enhanced by a tilted viewing angle). The branching in the GR seems to be hinting at multiple, overlapping ring features.

The GR consists of a thin, filamentary northern region, a clustered, ambiguous southern region (including the members of the GA), and filamentary branching towards the LHS.
Statistical assessment with elliptical shells, and optimum elliptical-shell-matching, identified two $> 4\sigma$ ellipse features corresponding to the GR prediction and to the visually-identified GR. Additionally, the 2D Power Spectrum Analysis (2D PSA) identified significant ($3.5 \sigma$) clustering on scales $\sim 320$~Mpc.

We also applied our statistical assessments to random data and to FLAMINGO-10K simulated data. The results demonstrate that, while superficially `significant' elliptical shells can be reproduced in random data with the optimum ellipse-matching method (many trials giving the `look-elsewhere' effect), with 2D PSA all of the random fields, and FLAMINGO-10K fields, were found to be entirely consistent with random. 

}

\begin{document}
\maketitle
\flushbottom

\section{Introduction}
\label{sec:intro}

The study of cosmological large-scale structure (LSS) is concerned with the distribution of matter in the Universe on the largest scales (i.e., at $\gtrsim 10^2$~Mpc, beyond the routine cosmic web). At these scales, structure is not expected to be gravitationally bound and can provide insights into the early Universe. The standard cosmological model ($\Lambda$CDM) explains the late-time structure, the cosmic web, as arising first from inflation and density fluctuations in the early  Universe and then cold dark matter (CDM) driving gravitational collapse along dense filaments and walls.
In this case, it becomes important for cosmology to understand the large-scale structure of the matter distribution in the Universe at late times as it may provide a window into understanding the early-Universe conditions. 

Ultra-Large Scale Structures (uLSSs) are apparently anomalous LSSs with physical dimensions exceeding that of the commonly-cited Yadav \cite{Yadav2010} scale of homogeneity: $\sim 370$~Mpc, proper size, present epoch.
Claims have been made of such physical structures, but they are often refuted by comparisons with simulations or simple statistical techniques. 
On the contrary, these uLSSs are not simple, but understanding their existence might further develop our understanding of the Universe.
Arguments may pendulum back and forth, but the observational data should continue to be investigated while there exists no consensus on the significance (or insignificance) of individual candidate LSS / uLSS discoveries. 

In this paper we present the discovery of the \emph{Giant Ring on the Sky}. 
This is the third paper of this kind; see \emph{A Giant Arc on the Sky} \citep{Lopez2022}, and \emph{A Big Ring on the Sky} \cite{Lopez2024}.
As previously, we are using the technique of mapping intervening Mg~{\sc II} absorbers present in the spectra of bright, background quasars to indicate the presence of low-luminosity, low-ionised gas which traces galaxies and galaxy clusters. 
The method has been successful in identifying intriguing LSS / uLSS (i.e., the Giant Arc and Big Ring) which might otherwise go missed with standard observational techniques. 
For more detail on this method, refer to the introductions in the original papers.

We include at the end of this section a short review of the Giant Arc (GA) and Big Ring (BR) in section~\ref{subsec:GA_BR_recap} for the reader's clarity in this newest discovery paper. 
The Giant Ring (GR) directly follows from the previous discoveries, so it is important that the reader be familiar with them.

\subsection{What is a `structure'?}

It is worth briefly addressing here what is meant by `structure' in the context of cosmological large-scale structure and ultra-large-scale structure. 
It is sometimes assumed that the term `structure'  is synonymous with `gravitationally-bound'. 
Conversely, many LSSs in cosmology, including superclusters (or superstructures) are \emph{not} gravitationally bound.

For a working idea of what might define a LSS (or uLSS) it seems reasonable to list some common qualities among claimed LSSs, as follows. 
(1) They are visually obvious and / or appear contiguous.
(2) They are often defined algorithmically through use of some type of friends-of-friends (FoF) algorithm, although not always. 
(3) They are overdense; in particular, their overdensities can approach or exceed $\delta\rho / \rho \sim 1$. 
(4) They have statistically-significant deviation from random expectations. 
(5) They might be traced by more than one independent tracer, if adequate data are available.
 
On a related matter, there might then arise the question of what differentiates a structure from a mere pattern in noise.
This particular concern follows from a recent claim of Sawala et al. \cite{Sawala2025} that the Giant Arc is simply a pattern in noise, and such GA-analogues can be easily reproduced in FLAMINGO-10K cosmological simulations.
In their analysis they use a standard FoF algorithm to identify \emph{candidate structures} of subhaloes within the FLAMINGO-10K simulation with memberships and overdensities akin to the GA.
According to their paper, they find many GA-analogues in nearly every sampling they investigate with memberships and overdensities resembling or exceeding the GA.
This information is in fact very misleading.
While they do find `GA-analogues': some with memberships exceeding the real GA and some with overdensities exceeding the real GA, they fail to demonstrate a single GA-analogue with \emph{both a membership and an overdensity} exceeding the GA.
This is a very important caveat of their work, and leads to very opposite conclusions. 
To elaborate, they in fact show that it is very difficult to reproduce a structure as large and as overdense as the real GA.
Moreover, it was emphasised above that the authors identify \emph{candidate structures}; this again is another important caveat of their work.
A candidate structure, or its direct field, must be assessed statistically to measure the likelihood of reproducing such a structure within randomly-distributed spatial data.
Previously, we have used a variety of statistical tests for the Giant Arc and the Big Ring (recapped in section~\ref{subsec:GA_BR_recap}) which determine the statistical significance of these structures and the corresponding fields. 
Without a statistical assessment of a candidate structure it is not likely possible to differentiate a pattern from a structure.
(Although one immediate difference is that a real structure will be usually be visually obvious.)
Thus, the work of Sawala et al. was unable to refute the GA, and was unsuccessful in showing that such structures abound in a $\Lambda$CDM Universe.\footnote{
The shortfalls of the Sawala et al. work are not limited to the points raised in this paper. We invite the reader to critically assess the work of the GA and the Sawala et al. rebuttal for themselves. 
}

In our work we have presented not only the Giant Arc, but also the Big Ring, and in this paper, the Giant Ring, all of which reside in the same field-of-view and in the same redshift slice (i.e., we have not looked everywhere to find these statistical anomalies).

\subsection{The Giant Arc and the Big Ring: a short review}
\label{subsec:GA_BR_recap}

The Giant Ring that is presented in this paper is an ultra-large-scale structure (uLSS) mapped by intervening Mg~{\sc II} absorbers in the spectra of background quasars. This newest discovery follows from the two previous uLSSs that were discovered with the same method: the Giant Arc \cite{Lopez2022} and the Big Ring \cite{Lopez2024}. We include here a brief summary of the Giant Arc (GA) and the Big Ring (BR) to set the scene for what follows. \\

\noindent
{\bf The Giant Arc} \\
The GA is an arc-shaped, uLSS extending close to $1$~Gpc at a redshift of $z=0.802\pm 0.060$.
It was the first uLSS presented using the method of Mg~{\sc II} absorbers in the spectra of quasars, and it added to an accumulating list of uLSSs that potentially challenge the assumption of homogeneity in the CP.
The GA was noted to have a resemblance to the Sloan Great Wall \citep[SGW,][]{Gott2005}, but larger and more distant, which led us to tentatively suggest that the GA could be a precursor to a structure such as the SGW.

The first hints of the GA became apparent when testing the method of Mg~{\sc II} tracing LSS \cite{Lopez2019}.
The GA was spotted visually, so the statistical assessment was unavoidably post-hoc. 
We performed a variety of statistical tests to assess the GA, these were: the Single-Linkage Hierarchical Clustering (SLHC) and the Convex Hull of Member Spheres (CHMS); the Cuzick and Edwards (CE) test; and the Power Spectrum Analysis (PSA). 
Each test assessed a different aspect of the GA and GA field, so one was advised to consider the evidence from the ensemble. 

The SLHC algorithm is equivalent to the Minimal Spanning Tree when separated at a specified linkage scale.
Such tests are fairly common in large-scale structure studies as they can be used to determine candidate structures in a field.
Note, in the specific case of the data that we use, the detection of a Mg~{\sc II} absorber is dependent on the availability of a background quasar at that exact location; the application of any spatial clustering (or otherwise) algorithm that does not account for these first-order inhomogeneities in the background probes, superimposed onto the inhomogeneities of the Mg~{\sc II} absorbers, should be used with caution. 
The SLHC algorithm, when applied to the GA field, identified the Giant Arc in two parts, which we named GA-main and GA-sub.
Then, using the method of the Convex Hull of Member Spheres \cite{Clowes2012}, the volumes encompassing the GA-main and the GA-sub were estimated.
These volumes were compared with random expectation volumes of the same number of points scattered at the control density to calculate the significance of the candidate structures. 
This method also supports the determination of the likelihood that the structure is real, and not just a pattern of noise in random data.
GA-main, containing $44$ Mg~{\sc II} absorbers, had a significance of $(4.5 \pm 0.6)\sigma$. 
GA-sub, containing $11$ Mg~{\sc II} absorbers, had a significance of $(2.1 \pm 0.9)\sigma$. 
We had visually identified the GA as the combination of the GA-main and GA-sub, but the SLHC algorithm split the structure into one large, statistically-significant portion, and one small, statistically-insignificant portion. 
The splitting of the GA could be indicating two independent structures.
However, given the proximity of the two (GA-main and GA-sub had overlapping absorbers on the sky), and that they both had the same MST-overdensity, we reasoned that the two were likely belonging to the same structure, and that perhaps a gap in the probes led to the apparent splitting of the GA.

The CE test is a case-control $k$ nearest-neighbours algorithm that calculates the significance of the clustering in a field with respect to inhomogeneous sampling --- i.e., this test accounts for the varying availability of the probes. 
This test does not, however, have the ability to test the GA on its own, or assess the physical scale of clustering.
We applied a series of `zooms' on the GA field to determine the significance of primarily the GA in the GA field.
The test yielded a tentative $3\sigma$ significance for the clustering in the field, but it did not give a physical scale for this clustering. (The results are $p$-values for the case-control $k$ nearest-neighbours; we used the results for comparison with other fields unrelated to the GA field.)

The PSA test is a Fourier method of assessing the clustering in the field on a physical scale (unlike the CE test). We applied the 2D PSA to the final `zoom' of the GA field which detected significant ($4.7 \sigma$) clustering on scales corresponding to $\lambda_{c} \sim 270$~Mpc.
We interpreted these results as the PSA detecting the clustering along the width of the GA.

Finally, the GA had some notable physical properties. (1) The SDSS DR16Q \citep{Lyke2020} \emph{field} quasars (i.e., those in the same redshift slice as the Mg~{\sc II} absorbers) appeared to be associated with the Mg~{\sc II} absorbers, providing independent corroboration. (2) We looked at the rest-frame equivalent width (EW) of the absorbers in the GA field and noticed a circle of strong absorbers in the centre of the GA which appeared to surround a small void. Additionally, the LHS of the GA had a higher concentration of strong absorbers versus the RHS. (3) We looked at the redshift distribution of the GA and again noticed an asymmetry along its length. The LHS of the GA was mostly concentrated in a small slice of the whole redshift slice located farthest from us, whereas the RHS of the GA appeared more diffuse across the whole redshift slice. \\

\noindent
{\bf The Big Ring} \\
The BR is a filamentary, ring-like uLSS extending up to $\sim400$~Mpc in its diameter.
It is again seen at a redshift of $z=0.802 \pm 0.060$ and is in the same field-of-view (FOV) as the GA.
The BR is the second uLSS discovered with the method of intervening Mg~{\sc II} absorbers.
On its own, it adds to the accumulating list of uLSSs that are potentially challenging the assumption of homogeneity in the CP.
However, given its proximity to the GA, it seems that these two features could be related in a way that might be difficult to explain within a $\Lambda$CDM Universe.

The BR first became apparent after switching from the older Zhu \& M\'enard \citep{Zhu2013} data to the newer Anand et al. \citep{Anand2021} (see section~\ref{sec:MgIIdata} below).
When we looked at the GA field with the new data, there appeared to be a curious ring feature to the north of the GA.
Again, this feature was spotted visually, so we inevitably performed our statistical analysis post-hoc. 
We repeated the use of the SLHC / CHMS method, and with an additional aspect of the Pilipenko Minimal Spanning Tree (MST) significance calculation \citep{Pilipenko2007}. 
We also repeated the CE test.
Finally, we experimented with a new method of objectively detecting filaments in the Mg~{\sc II} field with use of the FilFinder algorithm \cite{Koch2015}. The main results are summarised below. 

As was seen with the GA, the SLHC algorithm split the visually-identified BR, this time into 5 overlapping or adjacent groups.
We then applied the CHMS- and MST- significance tests to four different estimations of the BR for the final results, these were from visual (two estimates), SLHC, and FilFinder.
The additional use of the MST-significance was to counter the shortfalls of the CHMS calculation when applied to strongly-curved features.
The BR is a ring-like filament which is emphasised by the void-like regions around the ring and inside the ring. 
Therefore, the volume encompassing the BR is overestimated with a method like the CHMS.
The MST-significance calculation instead uses the mean MST edge-lengths between points, so it provides a better estimate for strongly-curved, filamentary features.
Using the CHMS-significance calculation, we obtained a significance of $(3.65 \pm 1.13\sigma)$.
Using the MST-significance calculation, we obtained a significance of $(4.10 \pm 0.45)\sigma$.
Notice that the CHMS-significance had a much larger error than the MST-significance, which is likely due to the volume overestimates.

Previously, when we applied the CE test to the GA field, we were able to use a method of zooming into the GA to determine, primarily, the signal of clustering from only the GA. 
Since the GA is a thin, filamentary arc, lying roughly horizontal on the tangent-plane images, we were able to reduce the $y$-axis significantly so that we were testing a small rectangular region containing primarily the GA.
We followed the same idea for the BR, but because the BR has a ring shape, we were restricted with our zooming in both axes.
The CE test did not find significant ($\geq 3 \sigma$) clustering in the field, as it had done with the GA.
However, a comparison of the $p$-value distribution against the case-control $k$ nearest-neighbours clustering scale, showed that the BR field had a similar clustering pattern to the GA field.
We also compared the BR field with four other fields that were unrelated and detached from the BR field.
These fields showed very arbitrary $p$-value distributions, and were very unlike the GA and BR field.

The FilFinder algorithm is an automated filament-identification tool that we applied to the Mg~{\sc II} image of the BR field. 
(We also applied the FilFinder algorithm to the SDSS DR16Q field quasars.)
Applying successively increasing size-threshold limits to the BR field demonstrated that the largest and most connected filament in the field belonged to the visually-identified BR (and an additional `Northern Spur' filament connected to the BR).

Finally, the BR had some notable observational properties.
(1) Similar to the GA, we found plausible association of the field quasars, and now also the DESI clusters \cite{Zou2021}, with the Mg~{\sc II} absorbers.
(2) We again looked at the distribution of EWs and found that there was a curved segment of low-redshift absorbers on the RHS of the BR with strong EWs. 
(3) Using a new project-plane method, we were able to view the absorbers in the BR at different angles. Doing so revealed two interesting properties of the BR. The first of these properties is that the BR appeared to have three distinct redshift bands, and from the side, the two outside bands made a curious `backwards S' shape. The second of these properties is that the BR also appeared to have a coil-like nature: from the front the BR is ring-like, but tilting the viewing angle revealed that the near-$z$ absorbers seem to `loop back' into the central ring feature. The central-$z$ ring contained the majority of the absorbers in a thin flat region, and the absorbers in the far-$z$ band appeared as a broken ring feature.

\section{Data and methods}
\label{sec:MgIIdata}

The Mg~{\sc II} data for this and our preceding papers depend on the detection
of the intervening Mg~{\sc II} absorption doublet ($\lambda\lambda$ 2796, 2803) in the spectra of quasars from data releases (DR) of the Sloan Digital Sky Survey (SDSS).

The discovery of the GA \cite{Lopez2022} arose from analysis of the Mg~{\sc II} catalogues of Zhu \& M\'enard \citep[Z\&M,][]{Zhu2013}
\footnote{See also https://www.guangtunbenzhu.com/jhu-sdss-metal-absorber-catalog.}, 
with their SDSS DR12 and DR7 `Trimmed' catalogues restricted to DR12Q \citep{Paris2017} and DR7QSO \citep{Schneider2010}, and with preference given to DR12Q in the case of repeated observations. 
The discovery of the BR \cite{Lopez2024} arose from analysis of the Mg~{\sc II} `QSO-based' catalogue of Anand et al. \cite{Anand2021}, with their SDSS DR16 data restricted to DR16Q \citep{Lyke2020}. For the current paper, we continue with the same (DR16Q-restricted) Anand et al. catalogue but make use also of the results of our own Mg~{\sc II}-finder applied to the DR16Q spectra. The intention here is to create a `high-confidence' sample that allows for the reliable detection of the doublet at signal-to-noise ratios that are a little lower than we used previously.

In constructing this high-confidence sample, the input data are the same --- the spectra corresponding to DR16Q --- but the selection algorithms are different, being those of Anand et al. and of our Mg~{\sc II}-finder. The expectation then is that adopting the doublets that are selected only by both algorithms will lead to few spurious detections, and hence allow us to have high confidence in the purity of the combined sample. Ideally too, the high-confidence sample will extend lower in
the signal-to-noise ratios of the doublets. Given the choice of continuing subsequently with the signal-to-noise ratios and other parameters calculated by Anand et al. or those  calculated by our Mg~{\sc II}-finder we have adopted ours (see section~\ref{subsec:adopted_parameters}).

To be sure that our high confidence is not misplaced, we checked visually some of the spectra, as detailed in section~\ref{subsec:visual_checks} below.

\subsection{Our Mg~{\sc II} finder}
\label{subsec:adopted_parameters}

Our Mg~{\sc II}-finder takes a ``heuristic'' approach --- that which seems to work. It is broadly similar to that of Raghunathan et al. \citep{Raghunathan2016}, but has been re-coded (in R) and incorporates some new features. The approach to continuum-fitting remains the same. We now smooth the input spectra so that detection of the doublet is more effective at low equivalent widths. The noise spectrum is generated by the same continuum-fitting approach, here applied to the absolute difference of the spectrum with respect to its continuum. Parameters for the detection of the doublet are initially set low so that they may be refined later, as required, by simple database operations, with no need to re-process all of the spectra. We use the parameters that we calculate for the doublets, rather than those of Anand et al., so that we have full flexibility. A particularly useful new parameter that we introduce is the continuum signal-to-noise ratio local to the doublets.

\subsection{Visual checks of the high-confidence sample}
\label{subsec:visual_checks}

We checked visually $100$ Mg~{\sc II} absorbers in the quasar spectra from the high-confidence Mg~{\sc II} sample. To ensure a broad representation of Mg~{\sc II} absorbers and quasar spectra, we looked at the spectra across  a range of $\lambda_{2796}$ redshifts ($z_{2796}$), $\lambda_{2796}$ `rest-frame' equivalent widths (EWs, or $W_{r, 2796}$), $\lambda_{2796}$ signal-to-noise ratios (SN), and quasar magnitudes ($i$). 

\begin{figure}
    \centering
    \includegraphics[width=\linewidth]{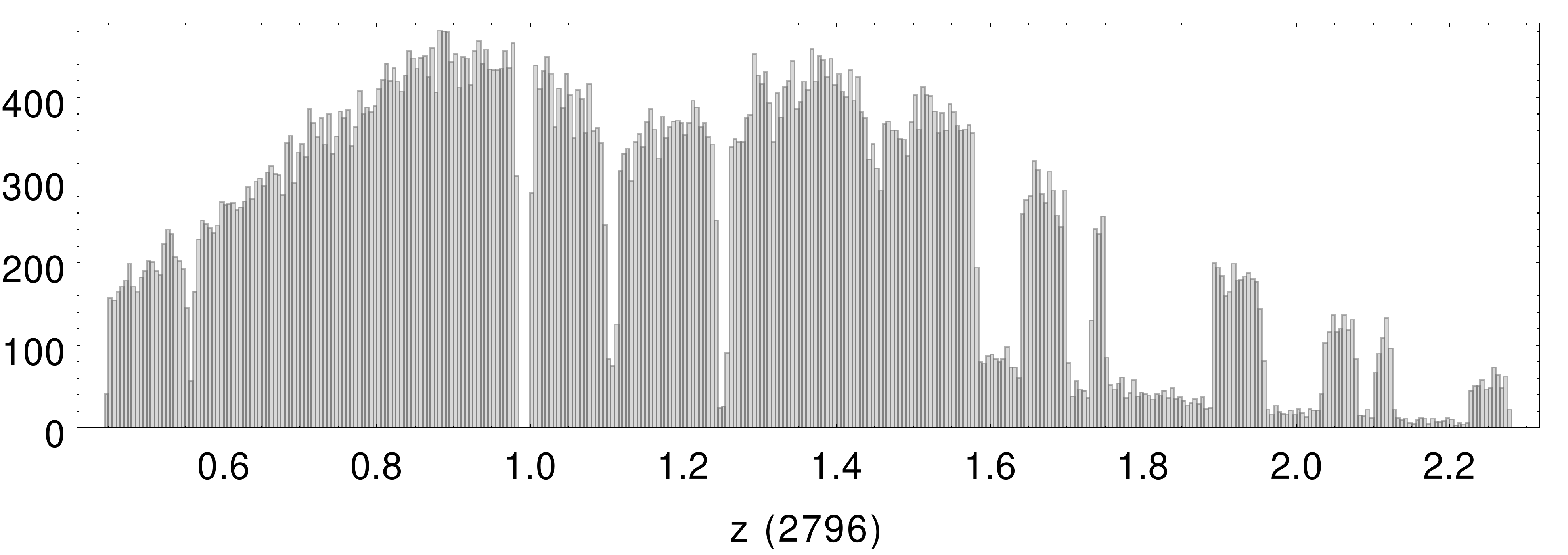}
    \caption{Histogram showing the redshift distribution of the $\lambda_{2796}$ Mg~{\sc II} line in the high-confidence catalogue of Mg~{\sc II} absorbers. There are noticeable dips in the distribution, for example, at redshifts $0.90$, $1.11$, $1.25$, and large drops at higher redshifts beyond $\sim 1.6$. These drops are typically a result of masking sky lines, such as the O~{\sc I} ($\lambda_{5577}$ and $\lambda_{6300}$), OH, and the high-pressure sodium bump at $\lambda_{5700}$.}
    \label{fig:high_confidence_z_dist}
\end{figure}

Five redshift bins were defined based on the $\lambda_{2796}$ redshift distribution of the high-confidence Mg~{\sc II} sample (see figure~\ref{fig:high_confidence_z_dist}): (1) $z_{2796} < 0.7$, (2) $0.7 \leq z_{2796} < 1.0$, (3) $1.0 \leq z_{2796}  < 1.3$, (4) $1.3 \leq z_{2796}  < 1.6$, and (5) $z_{2796} \geq 1.6$.

Within each redshift bin, $20$ quasar spectra were selected: one each for the minimum and maximum $W_{r, 2796}$, SN, and $i$; two spectra each at the mean $W_{r, 2796}$, SN, and $i$; one spectrum each at both $\pm 1 \sigma$ from the mean $W_{r, 2796}$, SN, and $i$; and lastly, two random spectra chosen by ordering the list by name and selecting two spectra roughly in the middle. Each of the $100$ spectra was visually inspected for the corresponding Mg~{\sc II} absorber.

We found that the full set of 100 visually-checked absorbers, before any additional cuts, had a $\sim 90$ per cent pass-rate for visually-convincing absorbers. The pass-rate per redshift bin was not uniform. Absorbers that were deemed inconclusive or false-positives were flagged. (Inconclusive absorbers could still be real of course, but visually we could not decide.) The least-successful bin was bin $1$ ($z < 0.7$), which had five flagged absorbers. Bins $2, 3$, and $5$ had two flagged absorbers each, and the fourth bin ($1.3 \leq z_{2796} < 1.6$) had a perfect pass-rate. 
Generally, we concluded that very high EWs are not always to be trusted: out of the five maximum $W_{r, 2796}$ absorbers (one per redshift bin), four of those were flagged; the exception was the absorber in bin $4$ which, perhaps not incidentally, had a much higher $\lambda_{2796}$ SN and brighter quasar magnitude than the other four. Other flagged absorbers had generally low $\lambda_{2796}$ SN values, but not strictly the minimum in each redshift bin. 

The high-confidence sample in its entirety pushes to lower SN values (Anand et al. values) of the doublet than the threshold cuts we imposed on the Anand et al. data for the previous work on the GA and BR. The above visual checks suggest that it is generally successful in doing so, and that threshold cuts might not now be necessary.
Nevertheless, threshold cuts (SN, $i$~magnitude) can still have some utility in: reducing contamination from remaining spurious absorbers; and diluting the
contamination of LSS features from unrelated absorbers (`LSS noise'), especially in a field with a large redshift-thickness.

\subsection{Project-plane method}
\label{subsec:projectplane}
Before continuing onto the next section, it is important for the reader to be familiar with the project-plane method.
The standard Mg~{\sc II} images that we show are 2D projections of 3D data (RA, Dec, $z$).
For the typical redshift slices that we look at, the physical redshift thickness has proper-size present-epoch measurement on the order of $10^2$~Mpc.
Consequently, any structuring along the line of sight (LOS) could be missed. 

Previously, when investigating the observational properties of the BR, we created a method for viewing different angles of 3D data as 2D projections: we call this the project-plane method.
The method allows for uncovering potentially interesting distributions of matter that could go missed when viewing the data from the standard LOS.
Additionally, we can utilise the project-plane method to investigate how different viewing angles enhance or reduce the visual impression or signal of potential structure candidates. 
The project-plane method is described in \citep{Lopez2024}, but here it is reiterated for the reader's benefit.

An initial, orthogonal, 3-vector coordinate system is defined such that $u_0, v_0, w_0$ are closely linked to $x_{prop}, y_{prop}, z_{prop}$. 
The initial normal vector ($w_0$) is defined as the proper coordinate to the mean $x_{prop}, y_{prop}, z_{prop}$ of the absorber members (which can be thought of as the original LOS). 
Then, the plane perpendicular to the normal is rotated such that $u_0$ points towards the most easterly absorber. 
Finally, all of the absorber members are projected onto the new plane.
Note that, with the project-plane method we are necessarily using the physical 3D distance coordinates associated with each data point, rather than the tangent-plane coordinates that are assigned at a specified redshift and FOV; in doing so we are also avoid large tangent-plane warping effects.
However, since the initial normal vector is usually closely linked to the LOS, the initial project-plane figures will have a similar projection to the standard tangent-plane Mg~{\sc II} figures, and their visual impression will be similar.

\section{A Giant Ring on the sky}

The first hints of a Giant Ring ultra-large large-scale structure (uLSS) became apparent when looking in the catalogues of Anand et al. in the field-of-view (FOV) containing the Big Ring (BR) and the Giant Arc (GA). 

\begin{figure} 
    \centering
    \includegraphics[width=0.7\linewidth]{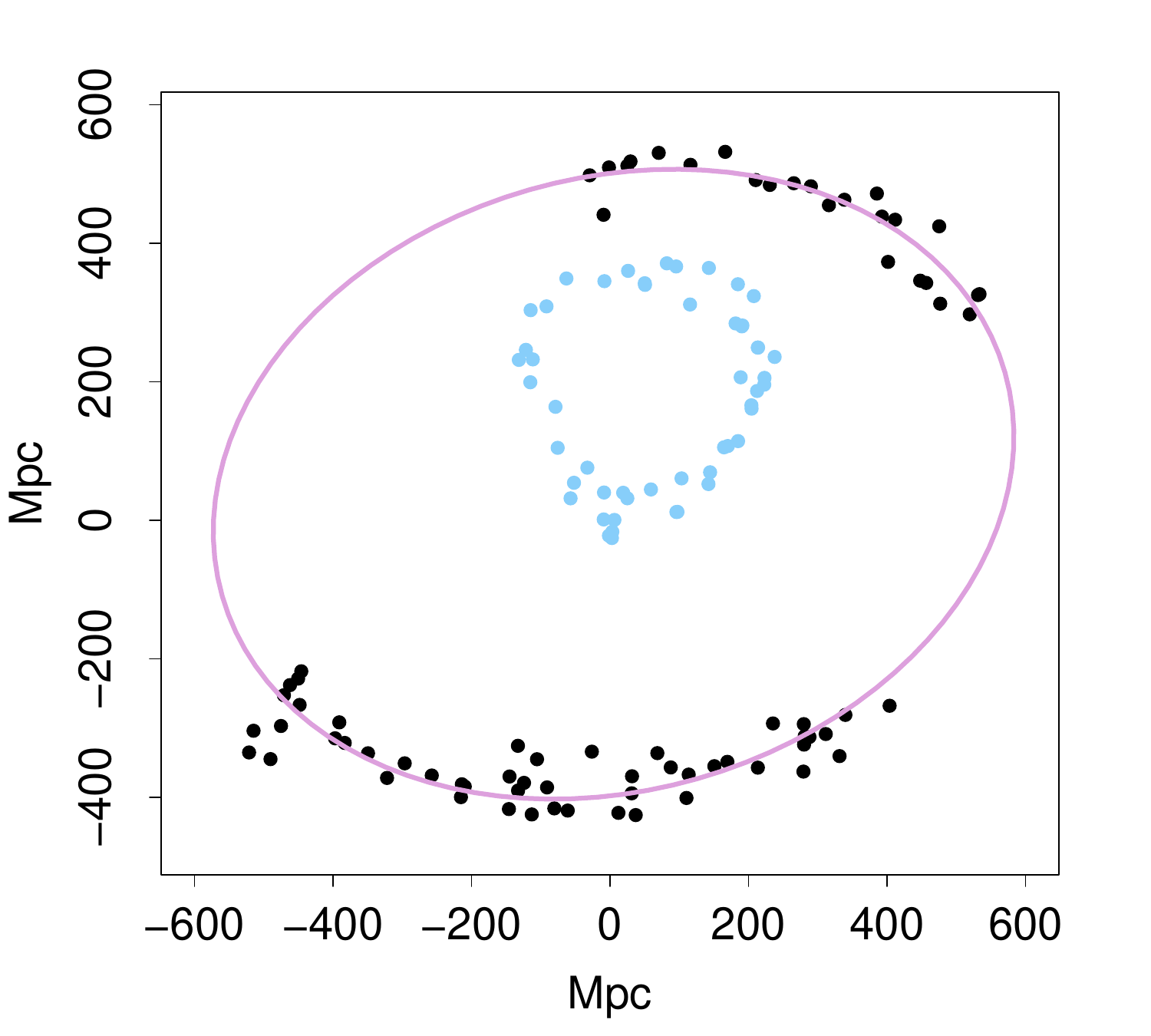}
    \caption{The Giant Ring (GR) prediction as indicated by the pink fitted ellipse joining the previously-identified Giant Arc (GA; black points, lower arc) and the new Northern Arc (NA; black points, upper arc); otherwise referred to as the GA+NA ellipse. The visually-identified Big Ring (BR) absorber points have been added with blue points. The GA+NA ellipse fully encompasses the BR from the approximate line-of-sight projection, creating the impression of two, not-quite concentric rings on the sky. The axes, which are the projected $u_0$ and $v_0$ vectors (see section~\ref{subsec:projectplane}), are in proper Mpc for the present epoch, where east is approximately towards the right and north is approximately towards the top, following the convention that $x$ increases towards the right and $y$ increases towards the top.}
    \label{fig:GR_original_prediction}
\end{figure}

In a review paper on investigating the GA and the BR together \citep{Lopez2025} we noted that the presence of a thin arc filament to the north of the BR, hereafter the Northern Arc (NA), appears to be an extension of the previously-identified GA. To demonstrate this point, we plotted the visually-identified GA absorber members (from the Z\&M catalogues), the visually-identified BR absorber members (from the Anand et al. catalogues), and the visually-identified NA absorber members (also from the Anand et al. catalogues) and added fitted ellipses.
The fitted ellipse joining the original GA and NA provided a provisional prediction of a potentially new uLSS, namely the Giant Ring (see figure~\ref{fig:GR_original_prediction}).

The discovery of a Giant Ring uLSS could have strong implications for cosmology. 
Consider the following questions that arise with the presence of a Giant Ring uLSS.
(1) The Giant Ring is, first and foremost, an uLSS discovery; its size exceeds the commonly-accepted Yadav scale of homogeneity \citep{Yadav2010} and may then indicate a potential challenge to the assumption of homogeneity in the Cosmological Principle 
(CP)\footnote{From a classical (historical) interpretation. We now know that the `scale of homogeneity' and the extent of homogeneity can have various interpretations; for a recent review, see Lopez et al. 2025 \cite{Lopez2025}. Nevertheless, it is still relevant to consider intriguing structures of vast scales, such as those exceeding the commonly-accepted scale of homogeneity, referred to as ultra-large-scale structures (uLSSs).}. 
(2) The huge size, and intriguing ring-like morphology becomes hard to explain within our standard cosmological model. 
Are uLSSs such as these seeded in the initial conditions? 
Since the GR, and other LSSs, are not expected to be gravitationally bound, what might be driving the formation of these structures, and how do we expect them to evolve? 
Do giant rings, and even nested rings, abound in a $\Lambda$CDM universe?
(3) The Giant Ring appears to be an extension of the previously-discovered Giant Arc, which would make this the second ring-like uLSS discovered with the method of intervening Mg~{\sc II} absorbers.
Both uLSSs, remarkably, are at the same redshift, in the same FOV, and with the larger ring appearing to encompass the smaller ring.
It now seems unlikely that a system of two almost-concentric,  ring-like structures could be the result of a look-elsewhere effect.
(The field of view containing both structures is now quite large, but we have so far analysed only one particular redshift slice; we have not yet looked elsewhere in the catalogue, unless checking for artefacts.)

Finally, we note here the remarkably similar redshifts of the Giant Ring reported in this paper, and the Giant Gamma-Ray Burst (GRB) Ring first reported in \cite{Balazs2015}.
The Giant GRB Ring is a huge ($\sim 1.7$~Gpc diameter), ring-like feature made up of nine GRBs in the redshift range $0.78 - 0.86$.
Similarly, the Giant Ring structure reported here is a huge ($\lesssim1$~Gpc diameter), ring-like feature made up $\sim 100$ Mg~{\sc II} absorbers in the redshift range $0.722 \leq z \leq 0.862$.
In the paper which first reports the Giant GRB Ring, the authors consider two interpretations of the ring: (i) that the orientation is such that the ring is nearly face on or (ii) the ring is a projection of a spheroidal structure.
Before having noticed the latter interpretation in their paper, we had begun to also consider the GR reported here as possibly a projection of a spheroidal structure.
It might therefore be worth bearing in mind that there now appear to be two, independent reports of huge, ring-like, structures around the redshift $z\sim0.8$, which could be interpreted as the projection of spheres.

\subsection{The GR as an extension of the GA}

The Giant Ring (GR) was predicted from the presence of a Northern Arc (NA) filament that appeared to extend the previously-known Giant Arc (GA); the GR prediction is otherwise referred to as the GA+NA ellipse.
However, in the Big Ring (BR) paper \cite{Lopez2024}, we noted that the GA looked somewhat different in the new catalogue of Mg~{\sc II} absorbers.
For the GA we had used the Mg~{\sc II} catalogues of Z\&M, corresponding to the SDSS quasar catalogues DR7QSO and DR12Q, and for the BR we began using the newer Mg~{\sc II} catalogues of Anand et al. corresponding to the SDSS DR16Q quasar catalogue (see section~\ref{sec:MgIIdata}).
The overall agreement between the selection of Mg~{\sc II} absorbers in the BR field for the probes in common was found to be $\sim 60$ per cent (see the BR paper for more detail).
However, the GA was still the most significant, most numerous and overdense structure detected in the field using the standard SLHC / CHMS methods described in section~\ref{subsec:GA_BR_recap}, despite the slight change in appearance.
Qualitatively, the GA appeared visually different in the new dataset --- the two likely reasons for this, as discussed in the BR paper, are reiterated here.
(1) The newer catalogues have many more quasar observations, and therefore corresponding Mg~{\sc II} absorber detections; the increased field density can alter the visual impression of the images. (2) There were several ($16$ out of $51$) GA absorbers missing in the Mg~{\sc II} catalogue of Anand et al. However, all Mg~{\sc II} absorbers belonging to the GA had been visually inspected and confirmed as real detections.

We are now using a high-confidence Clowes-Lopez Mg~{\sc II} catalogue (see catalogue description in section~\ref{sec:MgIIdata}).
The GA continues to appear visually different from the original GA in the Z\&M catalogue.
However, in this paper we demonstrate that, despite the qualitative differences, the GR is likely an extension of the GA; below we list a selection of findings which we believe to indicate this.
(Note, the following list is a preview of things to come later in this paper.)

\begin{itemize}
    \item[-] FilFinder identifies many absorbers closely related to the GR prediction (GA+NA ellipse), particularly the whole RHS of the GR.
    \item[-] FilFinder identifies branching on the LHS of the GR leading to an outer and an inner branch (which then rejoin at the GA). The outer branch appears closely related to the GR-predicted ellipse, whereas the inner branch appears closely related to the visually-identified GR (the estimate of the GR based on a visual impression by eye).
    \item[-] The largest, whole filament identified by FilFinder is the south-east quarter of the GR. When these identified points are superimposed with the original GA members the impression becomes that the GR is likely an extension of the GA (figure~\ref{fig:topcat_GR_FilFinder_GA} in section~\ref{subsec:FilFinder}).
\end{itemize}

We now speculate that overlapping ring features might be contributing to the ambiguity in the southern region of the GR.
By extension, we reason that the branching identified by the FilFinder algorithm (see section~\ref{subsec:FilFinder}) is also an indication of potentially overlapping ring features.
The outer branch of the GR appears more aligned with the GA+NA ellipse, whereas the inner branch of the GR appears more aligned with the visually-identified GR.

\subsection{Initial searches for the Giant Ring}
\label{subsec:initial_searches}
We are now using the high-confidence Clowes-Lopez Mg~{\sc II} catalogue; in this paper we investigate whether a GR is present in the data as indicated by the prediction from the original ellipse-fitting. 

\begin{figure} 
    \centering
    \includegraphics[width=1\linewidth]{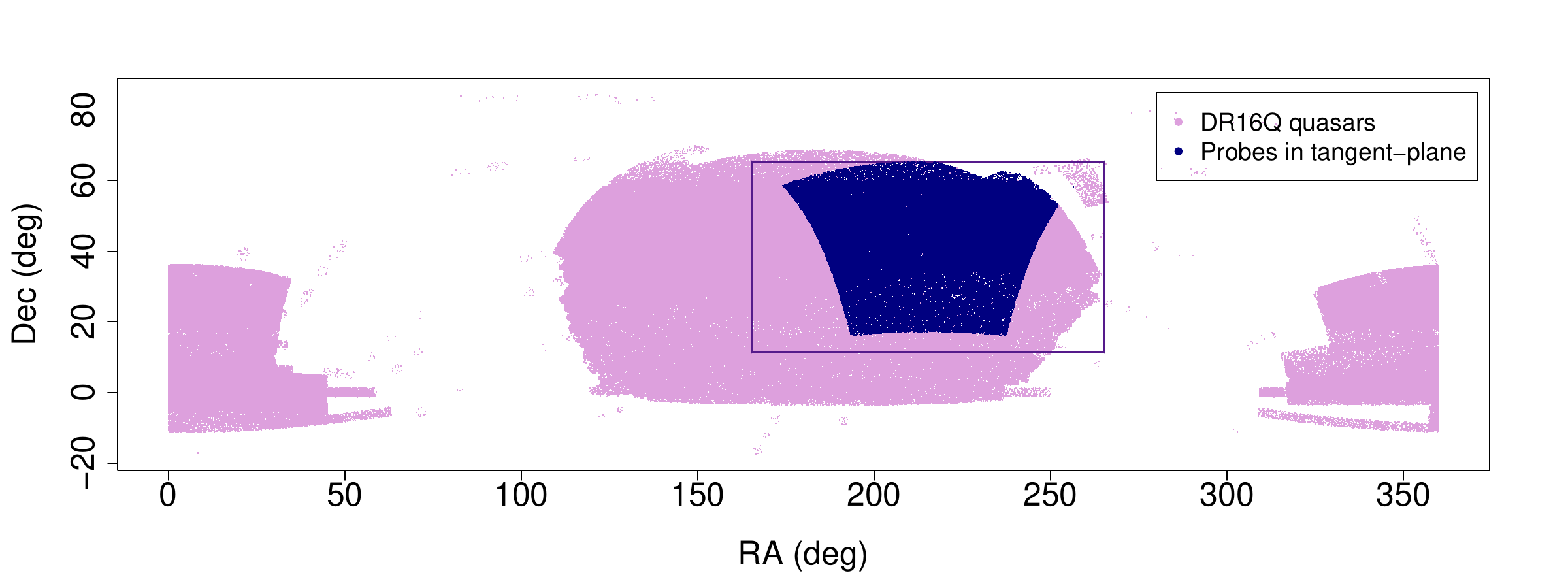}
    \caption{RA-Dec positions of the SDSS DR16Q quasars shown in light-pink points; these points also roughly mark out the SDSS DR16Q footprint. The purple, rectangular border indicates the initial RA-Dec selection for producing the Giant Ring images. In the process of defining a tangent-plane subset of points within this RA-Dec selection, some warping can occur, which is especially noticeable in large fields.
    The dark-blue points correspond to the tangent-plane subset, which are the points that are projected onto a rectangular plane which is tangential to the central (usually) RA, Dec, and $z$. The tangent-plane warping effects are obvious with the dark-blue points. The tangent-plane images are ideal for visual assessment, but in large fields the warping becomes problematic. For this reason, we use the project-plane images (section~\ref{subsec:projectplane}) where possible. } 
    \label{fig:prbs_tp_in_on_SDSS}
\end{figure}

Firstly, given the approximate size of the predicted GR ($\sim 1$~Gpc in diameter), we might expect that the GR is also vast in its redshift depth.
However, given the nature of the Mg~{\sc II} data and method, dealing with large redshift slices requires careful considerations (see section~\ref{subsec:redshift_slices} below). 
Secondly, again given the size of the predicted GR, we expect that the signal for the GR will be lower in comparison with a smaller, denser structure (such as the BR).
Experimenting with reducing the noise in the data should help to strengthen the signal of a real structure (conversely, for a structure that is the result of merely just random signals in noise, we would not likely find structure persistence when reducing the noise in the data). 
Finally, given the very large field-of-view (FOV) that we are now considering for the GR, our previously-used tangent-plane images become somewhat problematic due to the tangent-plane warping effects (see figure~\ref{fig:prbs_tp_in_on_SDSS}). 
We instead opt for the project-plane method where possible for producing the Mg~{\sc II} images for assessing and analysing the Mg~{\sc II} distribution (see section~\ref{subsec:projectplane}); however, in some cases, it is still appropriate to use the tangent-plane images. 

The initial searches for the Giant Ring consisted of a somewhat heuristic process of visually assessing the Mg~{\sc II} images with a series of different redshift slices and parameter selections. 
We were guided by the initial GR prediction and the above considerations of redshift size and the signal strength.
Conveniently, producing the Mg~{\sc II} images with the project-plane method also allows rotation and tilting of the viewing angle, which allows for the possibility of finding intriguing structures that do not happen to lie precisely orthogonal to the line-of-sight (LOS).

In our visual assessments we found that the GR became most pronounced upon widening the standard redshift slice ($z=0.802 \pm 0.06$) on the near-side by $\Delta z=0.02$ so that the new redshift range becomes $0.722 \leq z \leq 0.862$.
Additionally, we opted to reduce the noise in the data with signal-to-noise (SN) limits of $6, 3$~and~$12$ for the Mg~{\sc II} $\lambda\lambda 2796, 2803$ lines, and the local continuum estimation, respectively.
Note that, setting a \emph{local} continuum SN limit has some advantages over setting global continuum SN since the global continuum SN may not necessarily represent the exact continuum SN at the location of a Mg~{\sc II} doublet feature, and it is the doublet feature that we are primarily interested in measuring. 
Finally, we experimented with small tilts of the viewing angle of the Mg~{\sc II} images, which is conveniently quite simple within the project-plane method (see section~\ref{subsec:projectplane}). 
(Note, however, that large tilts might need the redshift range to be extended, because of truncated path-lengths at the extremities of the field.)

\begin{figure} 
    \centering
    \includegraphics[width=1\linewidth]{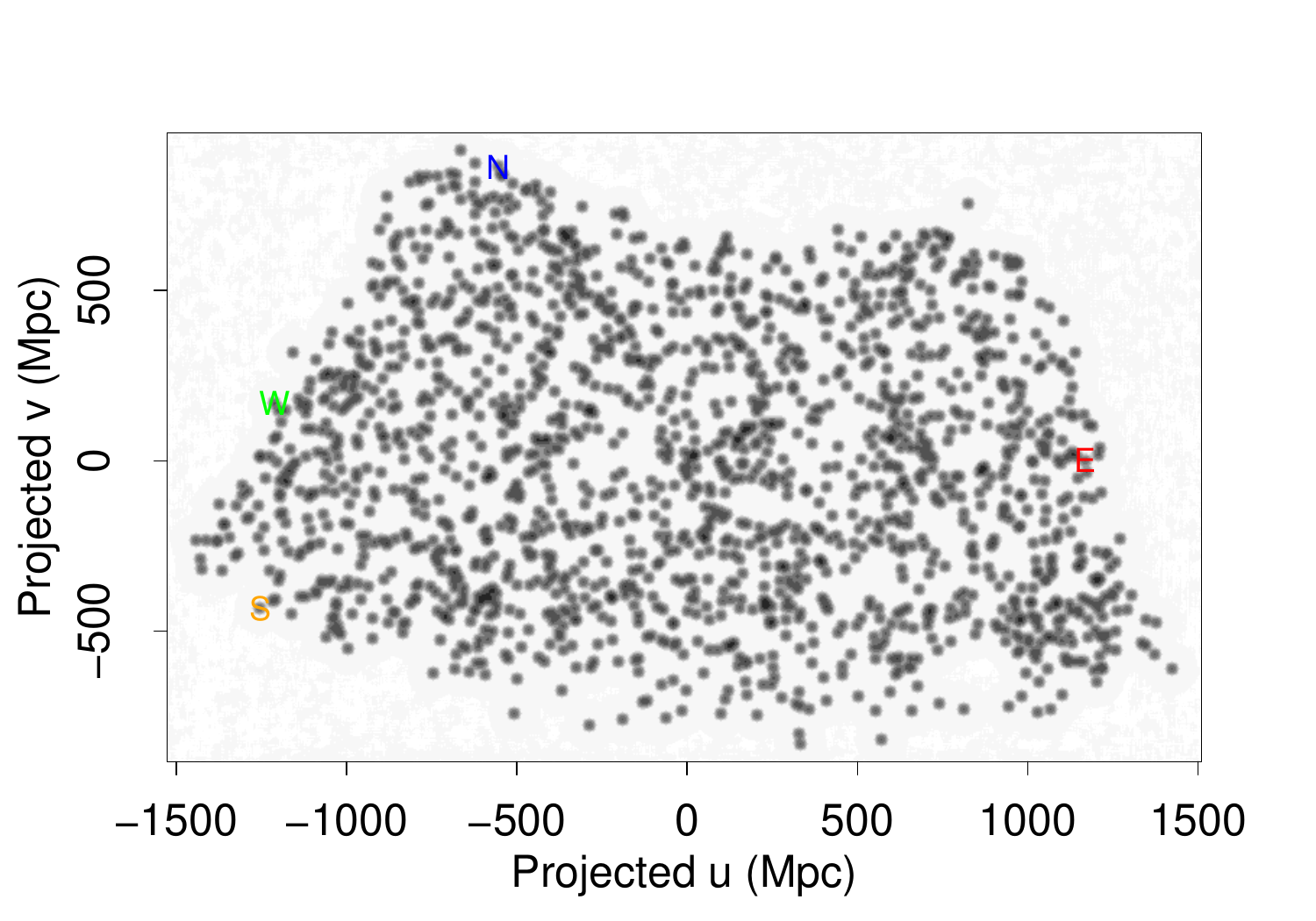}
    \caption{The project-plane distribution of Mg~{\sc II} absorbers in the redshift slice $0.722 \leq z \leq 0.862$. In this figure, the Mg~{\sc II} absorbers are projected onto the plane perpendicular to $w_0 + 0.1v_0$, which can be thought of as the original line-of-sight projection, modified by a small tilt. The axes, which are the projected $u$ and $v$ vectors, are in proper Mpc for the present epoch. The grey contours, increasing by a factor of two, represent the density distribution of the absorbers which have been smoothed using a Gaussian kernel of $\sigma = 9$~Mpc, and flat-fielded with respect to the mean Mg~{\sc II} distribution. Signal-to-noise (SN) limits of:  $6$, $3$ and $12$ were applied to the $\lambda_{2796}$,  $\lambda_{2803}$ Mg~{\sc II} lines and local quasar continuum, respectively (details of SN cuts are discussed in this section). The visually-identified Giant Ring (GR) can be seen around $x = -400, 600$ and $y=-500, 500$, i.e., a  roughly $1$~Gpc diameter, approximately circular ring. The GR is particularly emphasised by the void regions surrounding both the inside and outside of the ring annulus.}
\label{fig:GR}
\end{figure}

In section~\ref{subsec:projectplane} we briefly reiterated the project-plane method. 
With this method we can tilt and change the viewing angle of the data.
In a visual assessment of varying the tilt of the viewing angle, we found that tilting the $u_0, v_0$ vector-plane anti-clockwise with respect to the $w_0$ vector by $5^\circ$ brought a clearer image of the Giant Ring into view (see figure~\ref{fig:GR}).

For the rest of this paper, the standard GR field is according to the above selections, unless otherwise stated.
To reiterate, these are: within the redshift slice $0.722 \leq z \leq 0.862$ (but centred still on $z=0.802$ for producing tangent-plane images); SN limits of $6, 3$~and~$12$ for the Mg~{\sc II} $\lambda\lambda 2796, 2803$ lines, and the local continuum estimation; and tilting the viewing angle by $5^\circ$ using the project-plane coordinates. 
We also continue to use the images in the tangent-plane coordinates in some cases. 

\subsection{Initial checks for artefacts}
\label{subsec:initial_checks}
When using the Mg~{\sc II} absorbers as a tracer of the underlying LSS matter distribution, one needs to consider the combined effects of the quasar inhomogeneities (both intrinsic and arising from the survey) superimposed onto the inhomogeneities of the Mg~{\sc II} distribution itself.  
Unlike standard survey data, we are not observing or analysing bright objects directly, but instead our data requires a (ideally, uniform) sample of high-redshift, bright background probes to observe intervening matter, at a nearest redshift, at the RA / Dec location of the probes. 
Therefore, we must perform some standard checks to confirm that any LSS signal we detect in the Mg~{\sc II} absorbers is not an artefact of the probes. 
There are two ways in which we can test for this: 
(i) simply looking at the distribution of background probes and visually assessing whether there are inhomogeneities that correspond to the Mg~{\sc II} structure and 
(ii) check for repeating patterns in the Mg~{\sc II} absorbers that arise from the same set of background probes but in a nearer-redshift slice than the field of interest.
Both of these simple tests were used previously when analysing the Big Ring \cite{Lopez2024} and proved to be adequate.

\begin{figure} 
    \centering
    \includegraphics[width=1\linewidth]{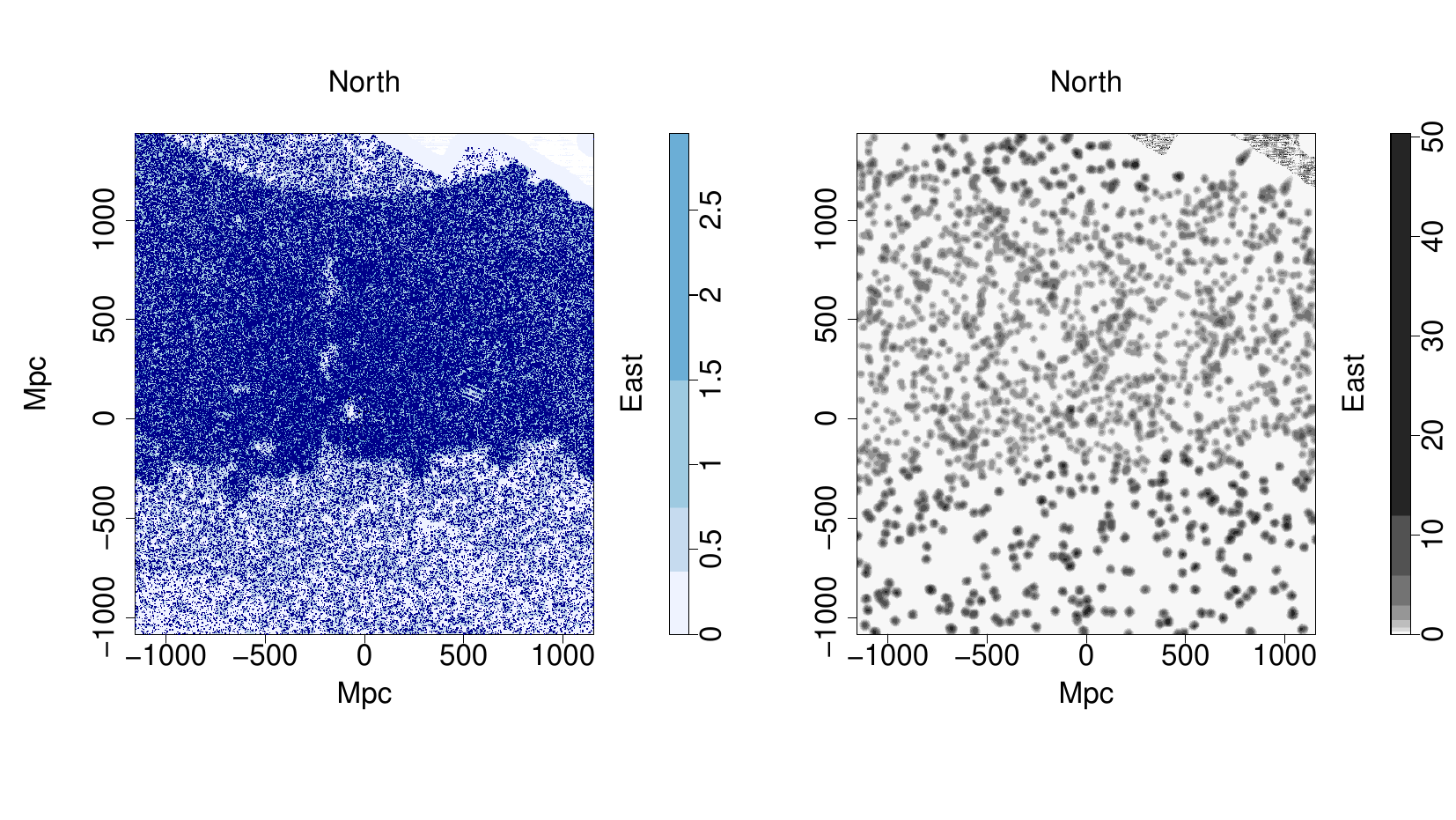}
    \caption{Tangent-plane distribution of background probes (left) and Mg~{\sc II} absorbers (right) in the Giant Ring field-of-view. The axes are in Mpc, scaled from the central Mg~{\sc II} redshift ($z=0.802$) to the present epoch. East is towards the right and north is towards the top, so that $x$ and $y$ increase towards the right and towards the top. Left: the background probes have the redshift condition $z>0.862$, i.e., farther than the far-edge of the Mg~{\sc II} redshift slice. Many inhomogeneities can be seen in this figure which are primarily from the survey (rather than intrinsic). In particular, there are harsh northern and southern borders where the density of quasars drops sharply. There are also patches of void-like artefacts of the surveys, which can be seen easily. Note that the curved border towards the top of the figure demonstrates the tangent-plane warping, which was discussed earlier in section~\ref{subsec:initial_searches} and shown in figure~\ref{fig:prbs_tp_in_on_SDSS}. Right: the corresponding Mg~{\sc II} absorbers in the redshift slice $0.722 \leq z \leq 0.862$ are here shown for the convenience of cross-comparing the density distribution. For the purpose of this assessment, there are no obvious artefacts that correspond to the dimensions of the GR in the distribution of background probes. } 
    \label{fig:GR_probes}
\end{figure}

The first check requires a visual assessment of the background probes. 
Figure \ref{fig:GR_probes} shows the tangent-plane distribution of the background probes corresponding to the GR Mg~{\sc II} redshift slice, i.e., quasars with the redshift condition $z>0.862$ in the GR FOV, and the tangent-plane distribution of the Mg~{\sc II} absorbers in the GR FOV on the right.
(Strictly, the far end of the Mg~{\sc II} redshift slice is at $z=0.866$, allowing for redshift errors of $0.004$, and strictly so too is the quasar redshift condition at $z>0.866$. However, we generally quote the redshifts which are set before the systematic addition of redshift-error allowance.) 
By blinking the two images together, we confirm that there are no obvious density artefacts that correspond to the dimensions of the GR.

\begin{figure} 
    \centering
    \includegraphics[width=1\linewidth]{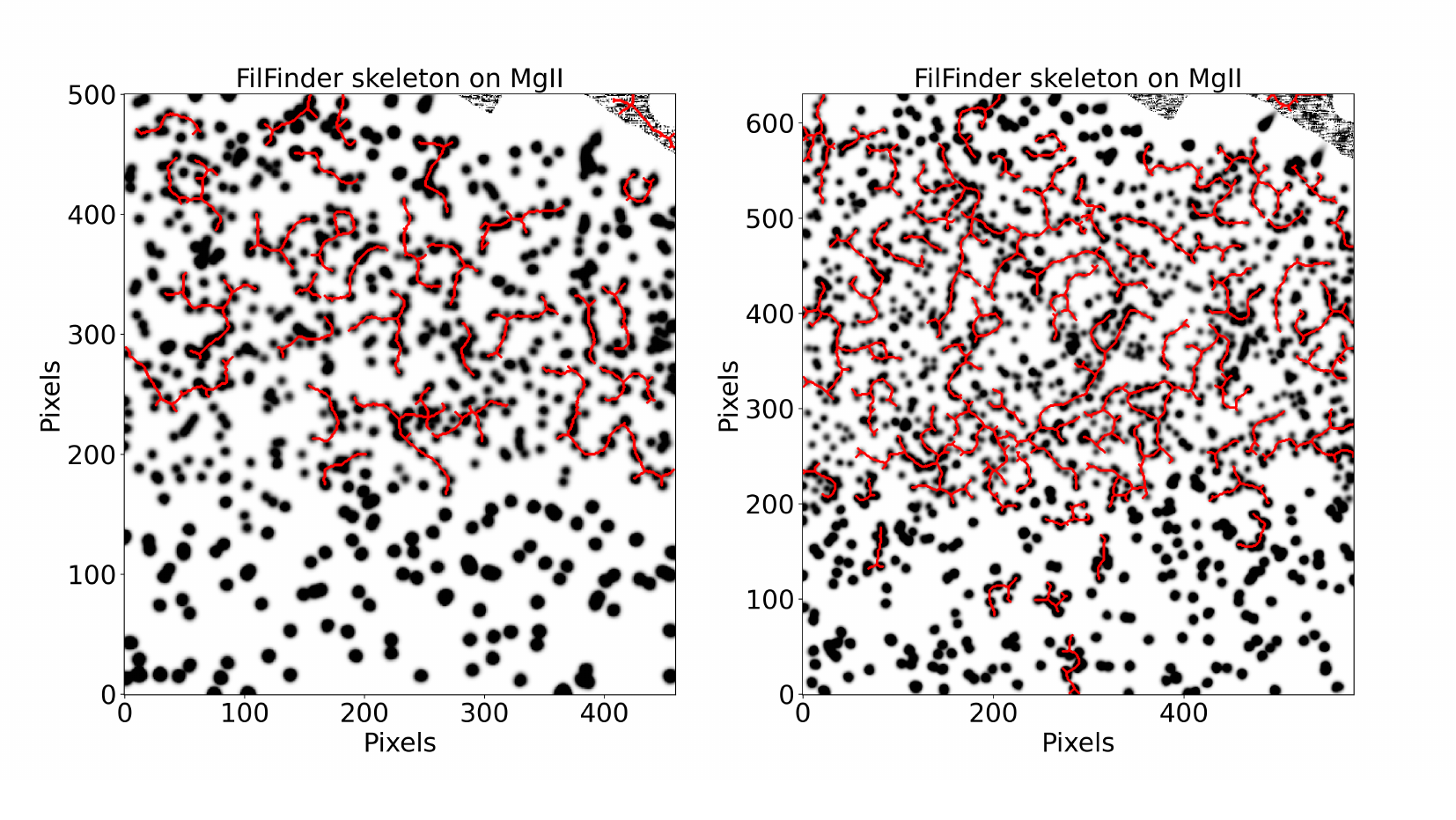}
    \caption{Skeleton results of the FilFinder algorithm applied to the Giant Ring (GR) and corresponding near-$z$ Mg~{\sc II} fields. The black pixels are the Mg~{\sc II} absorbers in the tangent-plane field which are smoothed (using a Gaussian kernel of $\sigma = 9$~Mpc) and flat-fielded with respect to the availability of background probes.
    The axes are in pixels corresponding to the proper-size, present-epoch Mpc scaled from the central Mg~{\sc II} redshift ($z=0.802$). East is towards the right and north is towards the top, so that $x$ and $y$ increase towards the right and towards the top. 
    The red lines are the FilFinder skeleton which highlight the filaments in the Mg~{\sc II} images.
    Left:  near-$z$ field ($z=0.600 \pm 0.060$) corresponding to the GR field-of-view (FOV) arising from the same set of probes as the GR field. Right: the GR FOV. Comparing the figures, we note that there are no obvious repeating filaments in the near-$z$ field corresponding to filaments in the GR field (with one potential exception of the lowest filament in the near-$z$ field, as discussed in the main text), further indicating that the GR (and BR also) is unlikely to be an artefact of the background probes.  }
    \label{fig:GR_nearz_fil}
\end{figure}

The second check requires a visual assessment of a near-$z$ slice that arises from the same set of probes as the GR field. In this assessment we are looking for obvious, repeating patterns in the Mg~{\sc II} distribution in the near-$z$ field that occur at the same location on the sky as the absorbers in the actual GR field, which might be indicating that artefacts in the probes are responsible for this Mg~{\sc II} distribution. 
As mentioned above, the GR arises in the same redshift slice as the previously-identified Giant Arc and Big Ring structures, but now we have extended the near-side of the redshift slice, i.e., $0.722 \leq z \leq 0.862$, with the central redshift at $z=0.802$ as usual. 
To be sure we are not looking at any potentially overlapping redshift slices in this particular assessment, we select Mg~{\sc II} absorbers in the near-$z$ field $z=0.600\pm0.060$ (arising from the same set of probes as the GR field, i.e., $z_{probe}>0.862$).
Again, with blinking, we can confirm that there are no obvious repeating patterns in the Mg~{\sc II} distribution in the near-$z$ field corresponding to the Mg~{\sc II} absorbers in the GR field
(see below and figure~\ref{fig:GR_nearz_fil} for a comparison with the FilFinder filaments).

For a more robust comparison of the GR field with the near-$z$ field, we can use the FilFinder algorithm \citep{Koch2015} (see also section~\ref{subsec:FilFinder}), which is an automated filament-identification algorithm. We have used FilFinder previously to identify filaments in the Mg~{\sc II} images.
With FilFinder, we can compare the filaments identified in each field by a standardised algorithm, rather than comparing, by eye, the Mg~{\sc II} density distribution.
Again, blinking the images of the FilFinder results on the GR field, and on the corresponding near-$z$ field, we find no obviously-aligned, repeating filaments. 
The only exception to this statement is the horizontal filament identified around the position  $x\sim200-300$ and $y\sim220$ --- this filament is near the lower part of the GR, and could be indicating an artefact in the probes here. 
However, a chance alignment is also plausible.
Nonetheless, we will proceed with caution regarding this particular section of the GR.
More importantly, however, note the difference in the visual impression given by both FilFinder images.
In the near-$z$ field the filaments appear random and small, but in the GR-$z$ field the filaments appear `swirly' and more connected.

With two different methods, we are able to show that the GR is unlikely to be an artefact of the background probes.
In the rest of this paper we will present the observational properties of the GR (section~\ref{sec:observational}), and the statistical analysis (section~\ref{sec:stats}). 
We also compare our observational and statistical results, where appropriate, with random Poisson data / expectation, and also with the FLAMINGO-10K subhalo data used in \cite{Sawala2025}, kindly supplied to us by Till Sawala.

\noindent
\subsection{A side note on redshift thickness} 
\label{subsec:redshift_slices}
Correctly dealing with redshift slices of real data, and of this specific type (bright background quasars as probes to the intervening, low-luminosity matter along the line-of-sight), is in fact rather complicated, and often over-simplified or misunderstood.

For example, in the work presented by Sawala et al. \cite{Sawala2025}
they point out that our handling of thin redshift slices inherently leads to large anisotropies. 
While it is true that the proper size of the redshift depth of the Mg~{\sc II} slices is inevitably smaller than the on-sky dimensions of the field\footnote{
For the GA work, the redshift slice had a proper-size thickness of $338$~Mpc, and the on-sky dimensions extended up to $\sim 2$~Gpc}, 
which will naturally lead to more anisotropic structures, it is also an unavoidable aspect of dealing with real data that are subject to redshifts, as well as data subject to the availability of background probes.
Consider the following two points.
(1) If one were to extend the redshift slice to have a thickness roughly equivalent to the on-sky dimensions ($\sim 2 $~Gpc), one would need to consider the appropriate interpretation of the redshift evolution within the slice. 
For a wide redshift interval, the evolutionary stage of the data at the far edge of the redshift slice becomes quite different from the evolutionary stage of the data at the near edge of the redshift slice.
So far, to our knowledge, simulated data has not accounted for this demonstration of redshift evolution. 
Simulated data is analysed at a single redshift,
whereas real data is observed and measured 
along the past light-cone. 
If one were to directly compare the two sets of data, then one would be over-simplifying the complexities of the real data (i.e., it is a case of comparing apples with oranges): attempts at drawing conclusions from such improper comparisons could be very misleading,
and the over-simplifications must be made very clear.
(2) As the redshift slice is increased, the number of available background probes decreases, which could lead to unnecessary loss of data for slices centred at low redshifts.
By reducing the data, we might lose the ability to detect intricate, low-signal, LSS. 
(Of course, for slices centred at high redshifts, the reduction of available background probes is unavoidable.)

In the case of the Giant Arc, we noticed an asymmetry in the spread of Mg~{\sc II} redshifts from the LHS to the RHS of the arc. 
As discussed in the Giant Arc recap above (section~\ref{subsec:GA_BR_recap}) 
the LHS of the GA was more compact, with the absorbers located mostly towards the far-side of the redshift slice, whereas the RHS of the GA appeared more diffuse across the whole redshift slice.
In this case, at least for the LHS of the GA, the thinness of the redshift slice does not seem to be the limitation, but rather the GA is (in reality) a long, thin, filamentary feature. 
Additionally, if in the case we were simply detecting
an artefact of an anisotropic redshift slice, then the dimensions of the feature would continue to correspond to the field. 
That is to say, there is no reason why the GA should have led to a Giant Ring (rather than a Giant Blob) of Mg~{\sc II} absorbers upon widening the FOV. 

An aspect of the work presented in this paper is the project-plane method. 
It is with this ability to tilt the viewing-angle of the field that we are able to view a visually-impressive, almost contiguous, roughly circular, Giant Ring feature.
The tilting might imply that the ring could have the morphology of a cylinder, which appears as a ring when viewed directly orthogonal to the circular bases. 
However, later in this paper we demonstrate that random spatial data with added spherical-shell clusters in a 3D box with dimensions matching the GR redshift can `re-create' ring-like features when viewed at different angles (see section~\ref{subsec:GR_analogues}).

Finally, careful judgment is required to interpret the data appropriately.
Some might prefer to dismiss such apparent anomalies as noise or artefacts.
However, we continue to accumulate evidence that these uLSSs might be pointing towards something interesting about our Universe.
Other than LSS and the CP, there are further anomalies in cosmology which appear to challenge the standard cosmological model (see \cite{Peebles2022} for a review, or \cite{Binney2024} for a manifesto of the Royal Society discussion meeting: \emph{Challenging the Standard Cosmological Model}); following where these hints lead might be fruitful for cosmology.

\section{Observational properties}
\label{sec:observational}

\subsection{2D nearest neighbours}
\label{subsec:2dnn}

The 2D nearest-neighbours method allows for the possibility of enhancing a low signal of clustering in a noisy background.
The GR project-plane images (in the $5^\circ$ tilted FOV) are plotted as usual, with an additional option to enhance the figures with a variety of 2D nearest-neighbour methods.
The options are: adding vectors of actual length, or fixed length, between each of the data points and its 2D nearest neighbour; adding vectors of actual length, or fixed length, joining pairs of 2D nearest neighbours, and isolating points which do not have a 2D nearest-neighbour pair within the specified maximum threshold; and adding vectors of actual length, or fixed length, joining pairs of 2D nearest neighbours, but removing entirely the points which do not have a 2D nearest-neighbour pair within the specified maximum threshold.
We define the fixed length and maximum threshold at $40$~Mpc.

With this method the idea is this: in a homogeneous sample, strongly-clustered points (small separations) are likely to represent the underlying structure, while the unclustered points likely contribute to the noise. 
Therefore, by removing the unclustered points, we may be able to enhance weak (but real) clustering signals.
The nearest-neighbours method can emphasise regions of higher (linear) density, analogous to Voronoi tessellation, in which the smaller Voronoi cells correspond to the higher densities.
Note that the work here is simple and qualitative (later we will present the statistical analysis of the GR).
\begin{figure}
    \centering
    \includegraphics[width=0.8\linewidth]{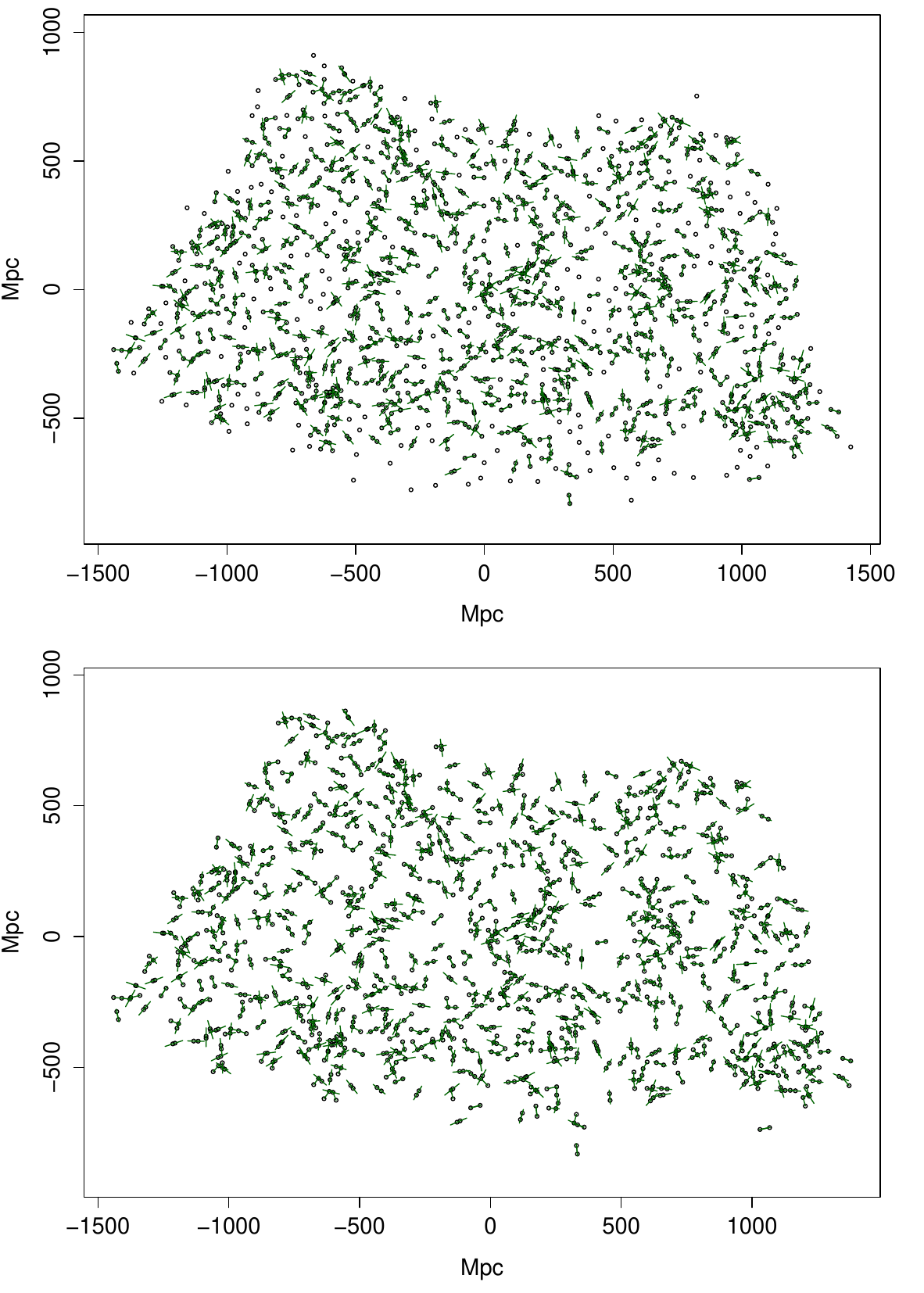}
    \caption{Two example figures using the 2D nearest-neighbour method. The figures show the tilted ($5^\circ$), project-plane GR field, as in figure~\ref{fig:GR}, but here with the Mg~{\sc II} absorbers plotted with small green points, and with added vectors between 2D nearest-neighbour pairs. 
    The axes, which are the projected $u$ and $v$ vectors, are in proper Mpc for the present epoch. 
    The fixed length of the vectors and the maximum threshold of the 2D nearest-neighbour separation are both set to $40$~Mpc. 
    Top: the added vectors of fixed length joins pairs of nearest neighbours, leaving isolated points which do not have a 2D nearest-neighbour pair within the specified maximum threshold. Bottom: the added vectors of fixed length joins pairs of nearest neighbours, but removing entirely the points which do not have a 2D nearest-neighbour pair within the specified maximum threshold (i.e., as above, but removing the isolated points entirely). In these figures, and especially the lower-panel figure, the density contrast between the void regions and the clustered regions is very marked. }
    \label{fig:2d_nn_plots}
\end{figure}

In figure~\ref{fig:2d_nn_plots} we show two figures of the GR field produced with the application of two different 2D nearest-neighbour methods.
Of note in the lower panel of figure~\ref{fig:2d_nn_plots} is the obvious void region towards the south-east of the BR which sits between the BR and the GR. 
In this figure, the density contrast between the clustered and void regions is very marked. Often in LSS studies the density contrast  ($\delta \rho / \rho$) is emphasised, but conceivably a relatively local measure of the \emph{contrast in the density contrast} between LSSs and their neighbouring voids could also be a useful parameter.

\subsection{FilFinder}
\label{subsec:FilFinder}

We saw the application of the FilFinder method in section~\ref{subsec:initial_checks} when we compared the GR field with its near-$z$ counterpart to show that the GR is not an artefact of the background probes. 
Here we briefly recapitulate the method (see also \cite{Lopez2024} and \cite{Lopez2025}) and then we discuss the FilFinder results on the GR field only.

The FilFinder method \cite{Koch2015} is an objective filament-identification method.
Its original application appears intended for use on small, filamentary, gaseous regions, such as star-formation regions.
However, the algorithm can be applied to any 2D pixel image, for which our data also complies.
During the analysis of the BR, applying the FilFinder method to cosmological large-scale structure was new, so we followed heuristic steps to determine what were appropriate parameter selections for our data. 
The parameter selections considered were: adaptive threshold, smoothing size, and size threshold;
we found that the most appropriate `standard' settings were ($18$, $12$, and $576$) pixels, respectively.

For the analysis of the BR, we next applied the FilFinder algorithm with incrementally increased size thresholds so that we could determine the largest and most connected filaments in the field. 
This process identified the BR in its entirety (plus an additional northern spur filament).
The GR is much larger and less concentrated than the BR, so the GR is made up of several large filaments;
we are here using the FilFinder algorithm to determine whether the visually-identified  and / or predicted GR is in agreement with the FilFinder-identified filaments.

In figure~\ref{fig:GR_nearz_fil} we saw the results of the FilFinder algorithm applied to the GR field.
The figure shows the GR in the standard tangent-plane coordinates (i.e., not the projected-plane view) where the Mg~{\sc II} parameter selections are as described in section~\ref{subsec:initial_searches} (but without the $5^\circ$ tilt, which is only applicable in the project-plane images).
Most notably, there are long filaments belonging to the GR which envelop the BR; these filaments are separated by large, obvious void-like regions.
The FilFinder results provide reasonable objective evidence that there is indeed a Giant Ring feature in the field, encompassing the previously-identified Big Ring.
The FilFinder results also demonstrate that the southern region of the GR is complex and ambiguous, whereas the northern region is very thin and filamentary.
It might also be noticed that the FilFinder algorithm identifies a similarly dense region of filaments to the left of the field which appear to extend from the ambiguous southern region.
The left-most collection of filaments create a thick, curved arc, which are again emphasised by the void regions between these filaments and the LHS of the GR.
We reason that it might even be possible that multiple overlapping ring features are responsible for the ambiguity in the southern region. 
Finally, note that this ambiguity in the southern region makes it difficult to determine the precise ellipse-fit of the GR.

\begin{figure} 
    \centering
    \includegraphics[width=1\linewidth]{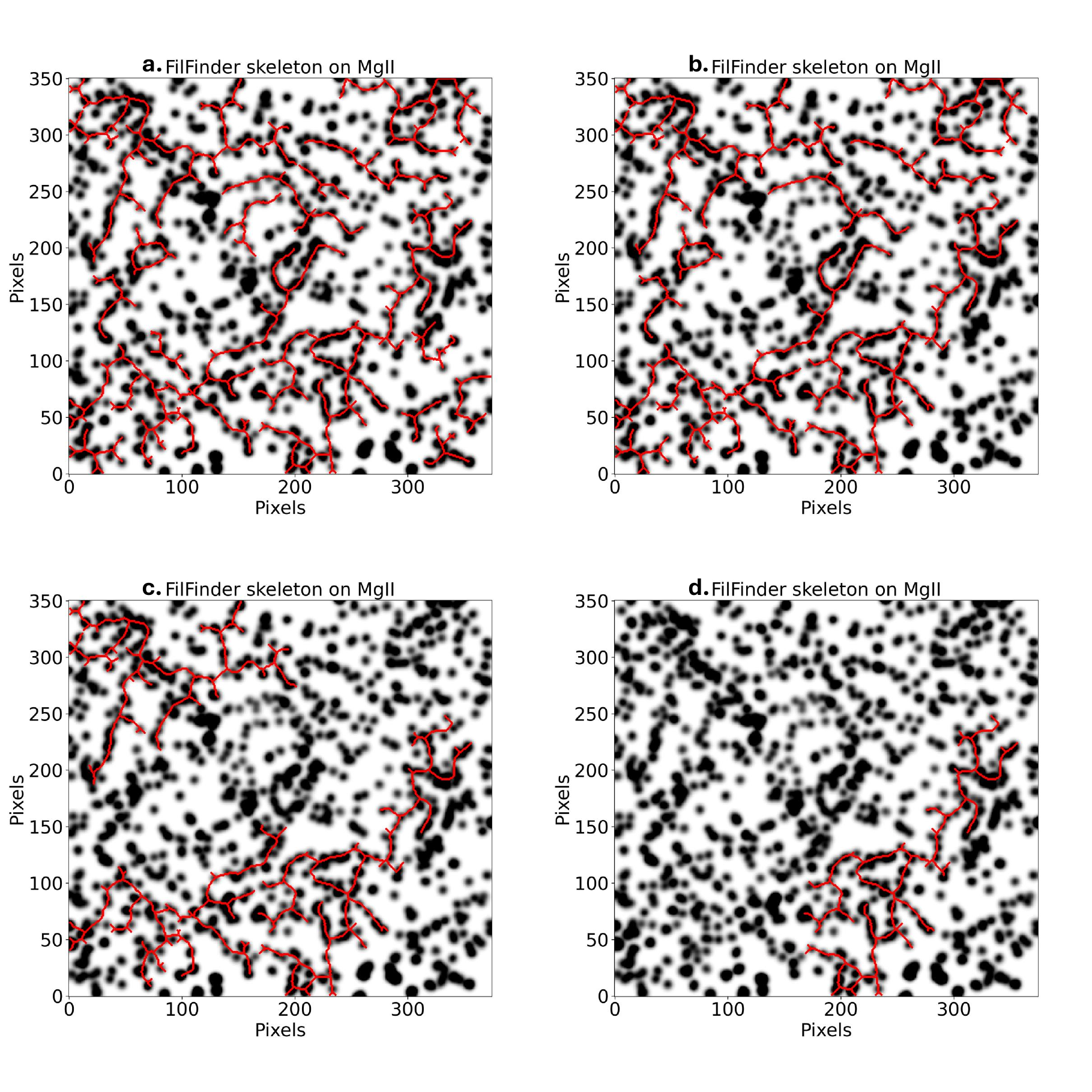}
    \caption{Skeleton results of the FilFinder algorithm applied to the tangent-plane distribution of Mg~{\sc II} absorbers in the Giant Ring (GR) field in a smaller field-of-view (FOV). The absorbers have been smoothed using a Gaussian kernel of $\sigma = 9$~Mpc and flat-fielded with respect to the distribution of background quasars. 
    The axes are in pixels corresponding to the proper-size, present-epoch Mpc scaled from the central Mg~{\sc II} redshift ($z=0.802$). East is towards the right and north is towards the top, so that $x$ and $y$ increase towards the right and towards the top. 
    The red lines are the FilFinder skeleton which highlight the filaments in the Mg~{\sc II} images. In each of the four panels the FilFinder algorithm is applied to the same field while successively increasing the size threshold parameter. Doing so allows the identification of the largest, densest, and/or most connected filaments.
    ({\bf a}) FilFinder applied with a size threshold $576$~pixels.
    ({\bf b}) FilFinder applied with a size threshold $1000$~pixels
    ({\bf c}) FilFinder applied with a size threshold $4000$~pixels.
    ({\bf d}) FilFinder applied with a size threshold $9000$~pixels. }
    \label{fig:FilFinder_GR_small_FOV}
\end{figure}

We can next look at a slightly reduced field, focussing on primarily the GR, and then successively increase the size threshold limit to find which are the largest, most dense, and connected filaments in the field (see figure~\ref{fig:FilFinder_GR_small_FOV}).
In each panel the size-threshold limit is: ({\bf a}) standard threshold $576$~pixels; ({\bf b}) $1000$~pixels; ({\bf c}) $4000$~pixels; and ({\bf d}) $9000$~pixels. (We did also check other intermediate threshold values, but we are here showing the four most distinct panels.)

\begin{figure} 
    \centering
    \includegraphics[width=0.8\linewidth]{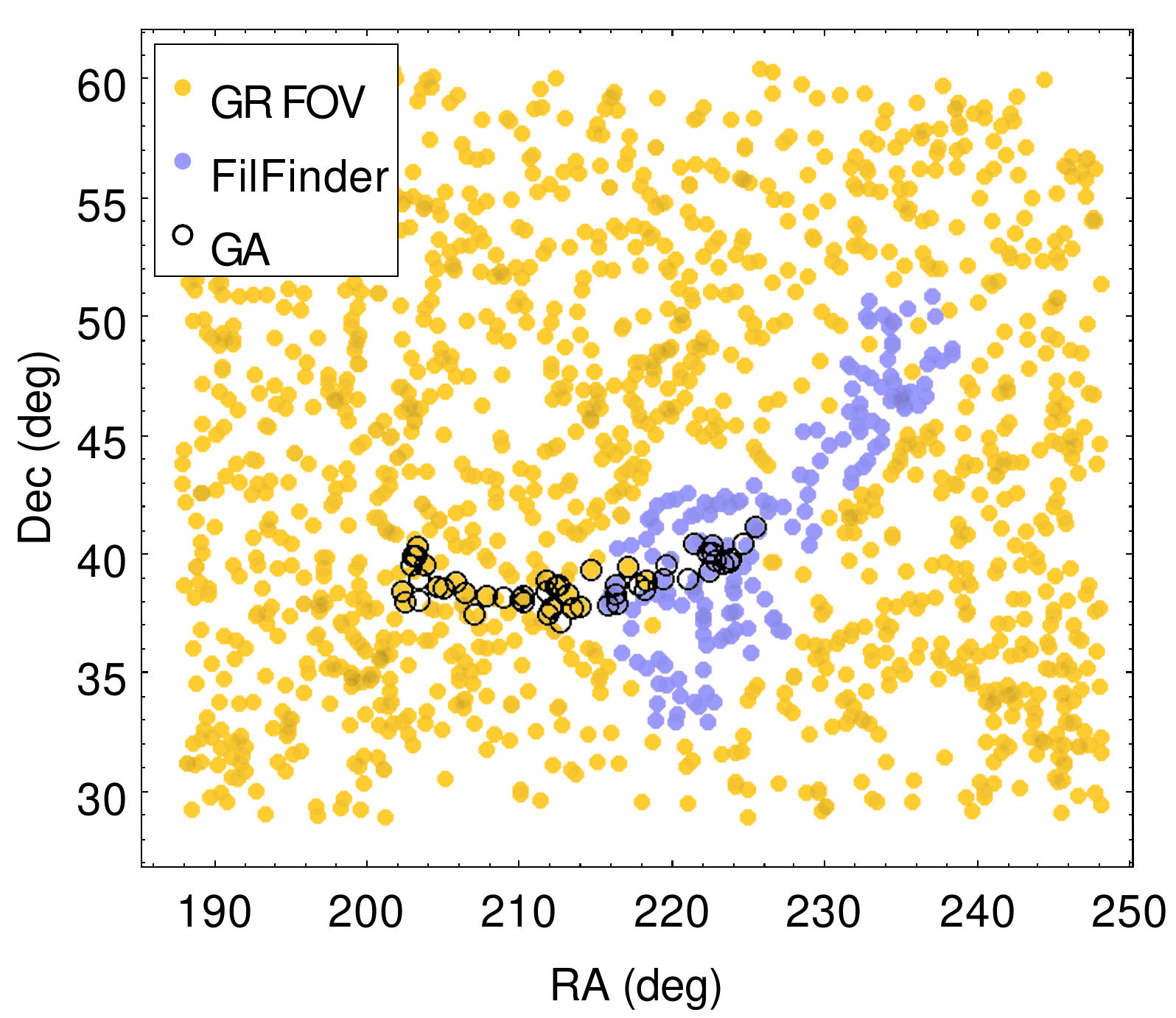}
    \caption{Topcat \cite{topcat} plot showing the RA-Dec positions of the Mg~{\sc II} absorbers in the small GR FOV (corresponding to the panels in figure~\ref{fig:FilFinder_GR_small_FOV}) shown in yellow. The lilac points correspond to the absorbers identified by the FilFinder algorithm with a size threshold limit of $9000$~pixels (panel {\bf d} in figure~\ref{fig:FilFinder_GR_small_FOV}). The open, black points correspond to the visually-identified GA (from the Z\&M Mg~{\sc II} catalogue). Note that the FilFinder-identified points were selected manually in Topcat.}
    \label{fig:topcat_GR_FilFinder_GA}
\end{figure}

Beyond a size threshold of $9000$~pixels there were no filaments remaining, and we saw no further breakdown of the large filament in panel {\bf d}.
If we then plot the FilFinder-identified absorbers in panel {\bf d} with the original Z\&M visually-identified Giant Arc members, we can see that the trajectory of the GA does indeed continue into this GR south-east filament (see figure~\ref{fig:topcat_GR_FilFinder_GA}). 
This south-east GR filament does however extend further south creating a more ambiguous region (i.e., the trajectory is not as clear). 
There is some indication that there might be overlapping features causing this ambiguity, which we mention again later in this paper.

If we now concentrate on panel {\bf a} we can see that the FilFinder identifies the visually-obvious GR. 
The north of the GR (the NA) is thin and filamentary, whereas the south of the GR (including the GA) is made up of a thicker region of clustered filaments.
If one were to trace the region of the GA, moving from right to left, then one would reach two diverging branches in the path; the left branch extends farther out, then rejoins the right branch towards the top of the GR, continuing into the NA section.
Interestingly, the branching appears to correspond to two slightly different versions of the GR: the outermost LHS branch closely corresponds to the GR as predicted from the GA+NA ellipse fit, and the innermost LHS branch closely corresponds to the GR that we visually identified in the tilted project-plane field\footnote{
Note: the FilFinder algorithm is applied to the tangent-plane projection of Mg~{\sc II} absorbers, whereas the visually-obvious GR was detected in the slightly tilted, projected-plane field. Since the tilt is only small, and the original projection closely resembles the line-of-sight, there are only minor shifts in the positions of the data points. This fact makes it fairly easy to judge the correspondence of absorbers between the different images.
}
--- we discuss the apparently two versions of the GR in more detail in section~\ref{sec:stats}.
The branching in the LHS of the GR, plus the ambiguity in the southern region of the GR (or GA) both suggest that there is a possibility of multiple, closely-aligned, overlapping ring features.

Inside the GR there are void regions separating it from the BR.
The impression is that of two not-quite concentric rings; a remarkable feature to observe in real data.
Apart from the GR and BR, there are two additional features identified by the FilFinder: a Northern Spur and a Tangential Bar.
The Northern Spur was first reported in the BR paper \citep{Lopez2024}.
It is the long, dense filament of absorbers extending from the BR towards the north-west direction.
The Tangential Bar is the diagonal filament pointing in the north-east direction that appears to connect the region of the GA / lower GR to the BR.
This feature was not present in the previous work; remember that we are here using a high-confidence sample of Mg~{\sc II} absorbers that combines the Anand et al. absorbers and our own catalogue, but parameters that we calculate for the doublets (which may allow for probing to lower EWs).  
Both features could be indicating interesting structure here, and will be noted for further investigation.

\subsection{Re-creating GR analogues}
\label{subsec:GR_analogues}
There is often disagreement on whether random data can reproduce the uLSS candidates that are reported in the literature.
For example, the Sloan Great Wall \cite{Gott2005}, the Huge Large Quasar Group \cite{Clowes2013}, and the Giant Arc \cite{Lopez2022}, were supposedly shown to be reproduced by random data or simulations \cite{Park2012, Nadathur2013, Sawala2025}.
The debate in LSS cosmology becomes a case of determining whether the reproduced LSSs in random or simulated data are representative of the real structure observed in the real data.
For instance, in the supposed GA-refutation from Sawala et al. \cite{Sawala2025}, they claim to easily reproduce the GA in simulations, but they do not assess the statistical-significance of their candidate structures. 
Judging only by the two parameters of the `GA-analogues' given, noting that not one of the listed `GA-analogues' had an overdensity \emph{and} a significance greater than or equal to those of the real Giant Arc, it seems unlikely that these candidate structures are statistically significant, and therefore also unlikely to be true analogues of the real Giant Arc.

Sylos Labini \cite{SylosLabini2026} pointed out that claims of massive structures being reproduced in simulations might actually arise from artefacts of the simulations, and that their `reality' would seem puzzling given the underlying physical
assumptions.
Furthermore, Di Valentino \cite{DiValentino2026} discusses the problem of `cherry-picking' results based on our own priors: for example, when data conform to what we expect, we consider them robust, and when data do not conform, we scrutinise the precision and accuracy. 
Indeed, in the specific case of LSS cosmology, if conclusions are made elsewhere that simulations can easily explain uLSSs results, then we might unjustifiably dismiss the real results more easily for our convenience.
However, we remain of the opinion that it is likely more productive for cosmology to question and investigate potential anomalous results rather than to casually dismiss them.
In the same way that we are sceptical of real data (systematics and biases), we should be sceptical of simulations (simplifications and assumptions).

We reiterate that the GA and BR were identified visually, and statistical analysis was unavoidably post-hoc.
By comparison, when attempting to reproduce the real features in random / simulated data, this visual aspect often goes missed, with the sole focus being on simple, perhaps naive, statistics --- often based on the outputs from FoF algorithms.
It is expected that FoF algorithms will find candidate structures in random data, but as we have emphasised previously, 
candidate structures should be assessed statistically and observationally, with evidence built from an ensemble of methods.
In the case of the structures we present, we provide multiple statistical and observational analyses.
Given that the visual aspect of assessing LSS is often under-appreciated, we have provided a very simple visual experiment in this section: we briefly demonstrate how one might reproduce `GR-analogues' in random data. 
\begin{figure}
    \centering
    \includegraphics[width=1\linewidth]{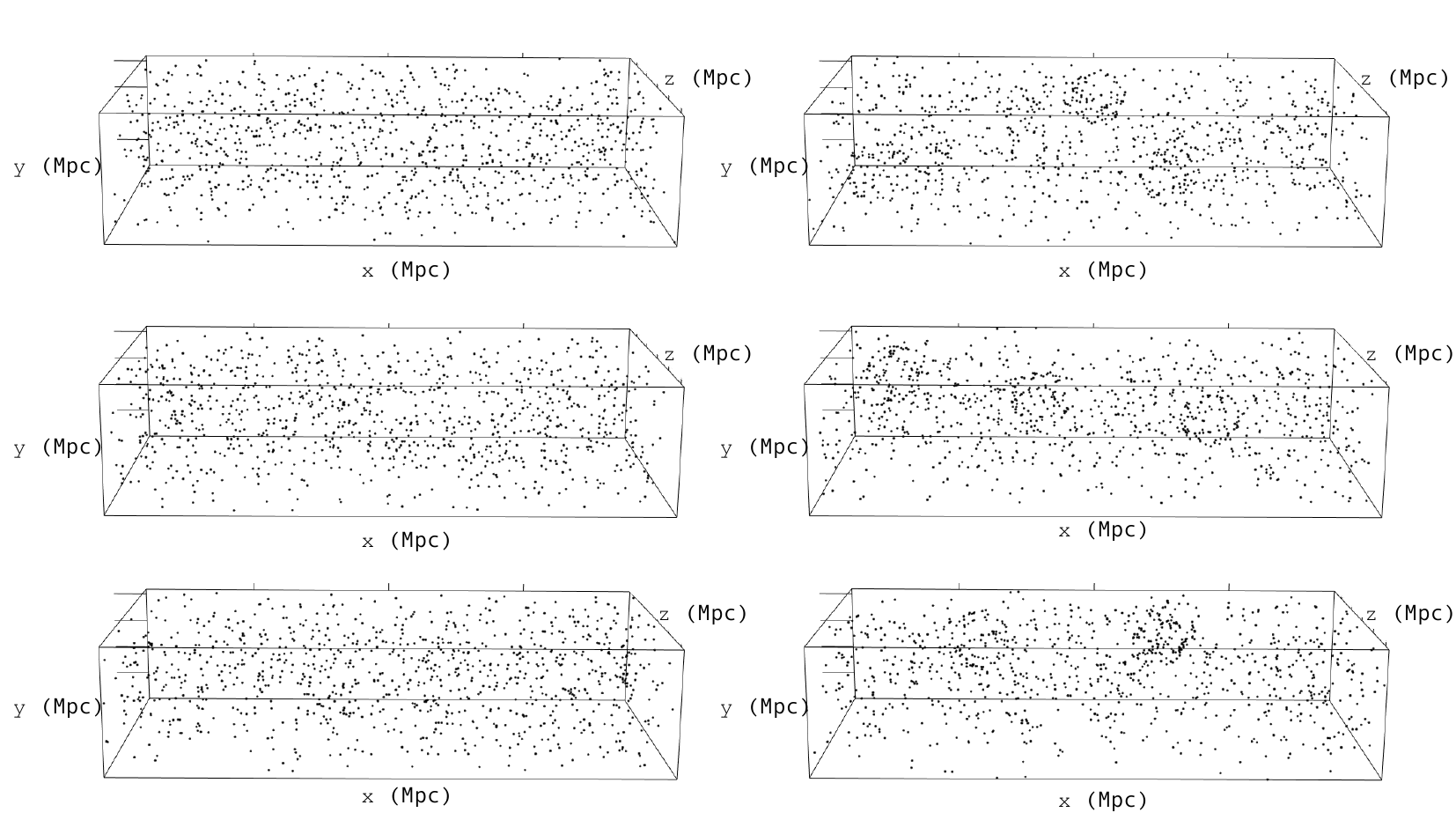}
    \caption{3D cuboids with dimensions  $(x, y, z) = (1800, 1000, 420)$~Mpc, mimicking that of the Mg~{\sc II} control field (see main text). In each cuboid, there are $964$ data points; in the three panels on the left, the points are distributed completely randomly (Poisson), and in the three panels on the right the data points are also distributed randomly, but with $150$ ($3 \times 50$) of the total points distributed within three spherical-shell clusters. The points within the (randomly located) clusters are distributed randomly within an inner radius of $95$~Mpc and an outer radius of $100$~Mpc.
    Visually, the panels on the left, containing just the random data, do not appear to have features comparable to the real data. By contrast, the panels on the right, containing the random data plus random spherical-shell clusters, appear to contain features similar to what we have observed in the real data. In particular, note the ring-like features, arcs, and even clumpy regions in the panels on the right.}
    \label{fig:random3D_and_clusters}
\end{figure}

A control field is defined with: $u= (-900, +900)$~Mpc, and $v=(-450, +550)$~Mpc. We are here using the project-plane field which has been tilted by $5^\circ$, containing the visually-obvious GR.
The redshift slice of the control data is $0.718 \lesssim z \lesssim 0.866$ (including an allowance of $\Delta z = 0.004$ for redshift errors at the near-side and far-side of the Mg~{\sc II} redshift slice), which corresponds to a present epoch, proper size of $420$~Mpc.
In this control sample, in the real data, there are $964$ Mg~{\sc II}  absorber points.
A 3D cuboid is then defined such that the dimensions of the three axes are equal to the dimensions of the control field in the real redshift slice containing the GR; that is, the cuboid has dimensions $(x, y, z) = (1800, 1000, 420)$~Mpc.
It is important here that the anisotropy of the redshift slice is adequately represented, i.e., the redshift thickness being significantly smaller than the on-sky dimensions.

A random sample of $964$ data points is distributed in the 3D cuboid. 
The points are plotted and shown within the cuboid --- see figure~\ref{fig:random3D_and_clusters}, left-hand side panels.
In general we find that, visually, the data here does not look like the real data that we observe with the Mg~{\sc II} absorbers.
Next, we then experiment with adding spherical-shell clusters within the random data.

Arbitrarily, we define three clusters, each with $50$ member points (taken from the total, so that the total number of points does not change), and with an inner radius of $95$~Mpc and an outer radius of $100$~Mpc; the $50$ points are distributed randomly within the specified radii, and the centre points of the clusters are also positioned randomly.
Contrastingly, the distribution of the random points plus spherical-shell clusters looks visually closer to what we see in the real data --- see figure~\ref{fig:random3D_and_clusters}, right-hand side panels.
In particular, note the ring-like features, arcs, and even clumpy regions.

Clearly, this visual experiment is designed only to make and illustrate a point.
(We later noted that Bal\'azs et. al. \cite{Balazs2015} made a similar case for spheroidal structure as a possible explanation for seeing a giant ring in the distribution of GRBs. In their analysis, they used Monte Carlo simulations to project a spheroidal shell onto a plane --- in many examples these structures appeared as rings.)
However, more generally what we demonstrate here is the possibility that what we observe in real data might be indicating additional inhomogeneities that random (or even simulated) data is not producing.
Later, we perform statistical analysis on random and simulated data and compare these results with the observed data.
In general, we find that a single candidate structure assessment is insufficient to present the whole story.
Instead, using multiple tests concerning both the individual candidates \emph{and} their surrounding fields indicates that the real data may be inhomogeneous or anomalous against expectations.

\section{Statistical Analysis}
\label{sec:stats}

Commonly in cosmological LSS analysis, 
Minimal Spanning Tree (MST) methods are used (otherwise known as Friends-of-Friends, or FoF) to identify candidate structures in the field. 
In our previous works of the GA and BR, we made use of 
the Single-Linkage Hierarchical-Clustering \citep[SLHC;][]{Clowes2012} algorithm, which we used together with the Convex Hull of Member Spheres \citep[CHMS;][]{Clowes2012} or alternatively the Pilipenko MST \cite{Pilipenko2007} for assessing the significance of the identified candidate structures (see also  section~\ref{subsec:GA_BR_recap}). 
In the papers, we emphasised that these MST methods are not strictly appropriate for our data, given the non-uniform distribution of background probes, even in reasonably homogeneous regions. 
The gaps across the quasar data inevitably contribute to the inhomogeneities in the Mg~{\sc II} in some complicated way --- disentangling the two would be onerous. 
In any statistical assessment of the Mg~{\sc II} data, we have to consider the complications that arise 
from the availability of background probes. 
However, for MST tests in particular, which rely on the separation between data points, these complications have a greater detriment. 
Given the above concerns, and to avoid the misinterpretation of the statistical results, we choose not to utilise the MST method for identifying or quantifying candidate structures.
Instead, we here describe a novel (in that our use of it is new) statistical method of assessing with elliptical shells (section~\ref{subsec:ring_tests}).
Additionally, we measure the field clustering with the Power Spectrum Analysis (section~\ref{subsec:PSA}).

\subsection{Assessing with elliptical shells}
\label{subsec:ring_tests}

The GR was predicted from fitting an ellipse to the original Z\&M GA and a newly-identified (in the Anand et al. data) Northern Arc (NA) filament --- the GA+NA ellipse.
Guided by this prediction, we later found a visually-impressive, almost contiguous, roughly circular, GR
in the slightly-extended redshift interval $0.722 \leq z \leq 0.862$, using SN parameters of $6, 3$ and $12$ for the Mg~{\sc II} $\lambda\lambda2796, 2803$ lines and the local continuum estimation, respectively, and with a viewing angle tilted by $\sim 5^\circ$ (see section~\ref{subsec:initial_searches}).
In this section we show the results of assessing with elliptical shells for the parameters of the predicted ellipse (in both the original and tilted projected planes), and later with optimised ellipses (section~\ref{subsubsec:optimum_ellipse}). 

The test described here is a simple assessment using elliptical shells. 
By defining two concentric ellipses separated by some small value $\Delta r$, we create an elliptical `shell' with a thickness $\Delta r$. 
Within the shell, the number of data points is counted and compared with the expectation value (for a Poisson distribution) and with random simulations.  
Later in the paper, the method will  be adapted to test a range of ellipse parameters, orientations, and positions, to optimise the ellipse-fitting (section~\ref{subsubsec:optimum_ellipse}). 

\begin{figure} 
    \centering
    \includegraphics[width=0.8\linewidth]{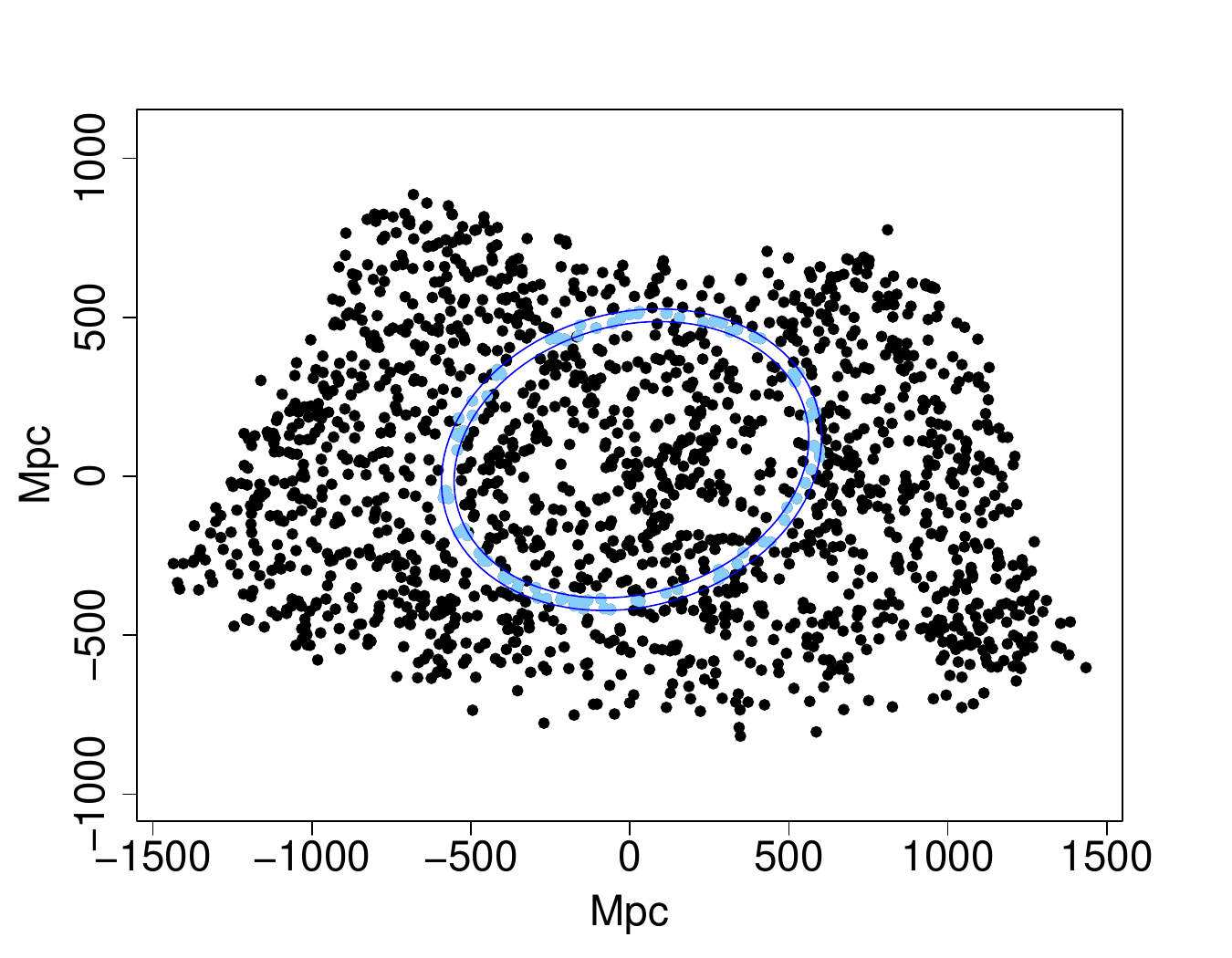}
    \caption{The project-plane distribution of Mg~{\sc II} absorbers in the GR field (standard viewing angle, no tilt) shown as solid, black points. The axes, which are the projected $u_0$ and $v_0$ vectors, are in proper Mpc for the present epoch. The blue elliptical shell corresponds to the GA+NA ellipse (the GR-predicted ellipse). The solid, blue points are the Mg~{\sc II} points that fall within the elliptical shell. Note that the GA+NA ellipse (the GR-predicted ellipse) is a little different from the visually-identified GR, as discussed throughout this paper. That is, the LHS of the predicted GR extends farther to the left, appearing to cross into a different overdense arc, which is adjacent to the innermost LHS branch belonging to the visually-identified GR --- the branching is discussed in section~\ref{subsec:FilFinder}. }
    \label{fig:GR_first_shell}
\end{figure}

In figure~\ref{fig:GR_original_prediction} we showed the predicted GR from the GA+NA ellipse (which encompasses the Big Ring).
The parameters of this ellipse in the standard projected plane are:
\[a = 588; b = 442; \epsilon=0.248; \theta = 0.278; (x, y) = (5.12, 52.2), \]
where $a$ is the semi-major axis, $b$ is the semi-minor axis, $\epsilon$ is the ellipticity of the ellipse (calculated as $\epsilon = 1-b/a$), $\theta$ is the angle of the ellipse anti-clockwise from horizontal in radians, and ($x, y$) is the centre of the ellipse in the standard project-plane field.
The semi-major and semi-minor axes of the central ellipse are then increased by $\Delta r /2$  to give an outer ellipse, and decreased by $\Delta r / 2$ to give an inner ellipse, which together form the elliptical shell with a thickness $\Delta r$ (see figure~\ref{fig:GR_first_shell}).
In these tests, we choose a thickness $\Delta r = 40$~Mpc since we are primarily interested in the thin, filamentary, GR feature. 
Choosing a $\Delta r$ that is too large will simply measure the expectation value, and choosing a $\Delta r$ that is too small will likely miss the primary features (since the ellipse-fitting assumes a perfect ellipse, and the data might not entirely comply). 

\begin{figure} 
    \centering
    \includegraphics[width=0.8\linewidth]{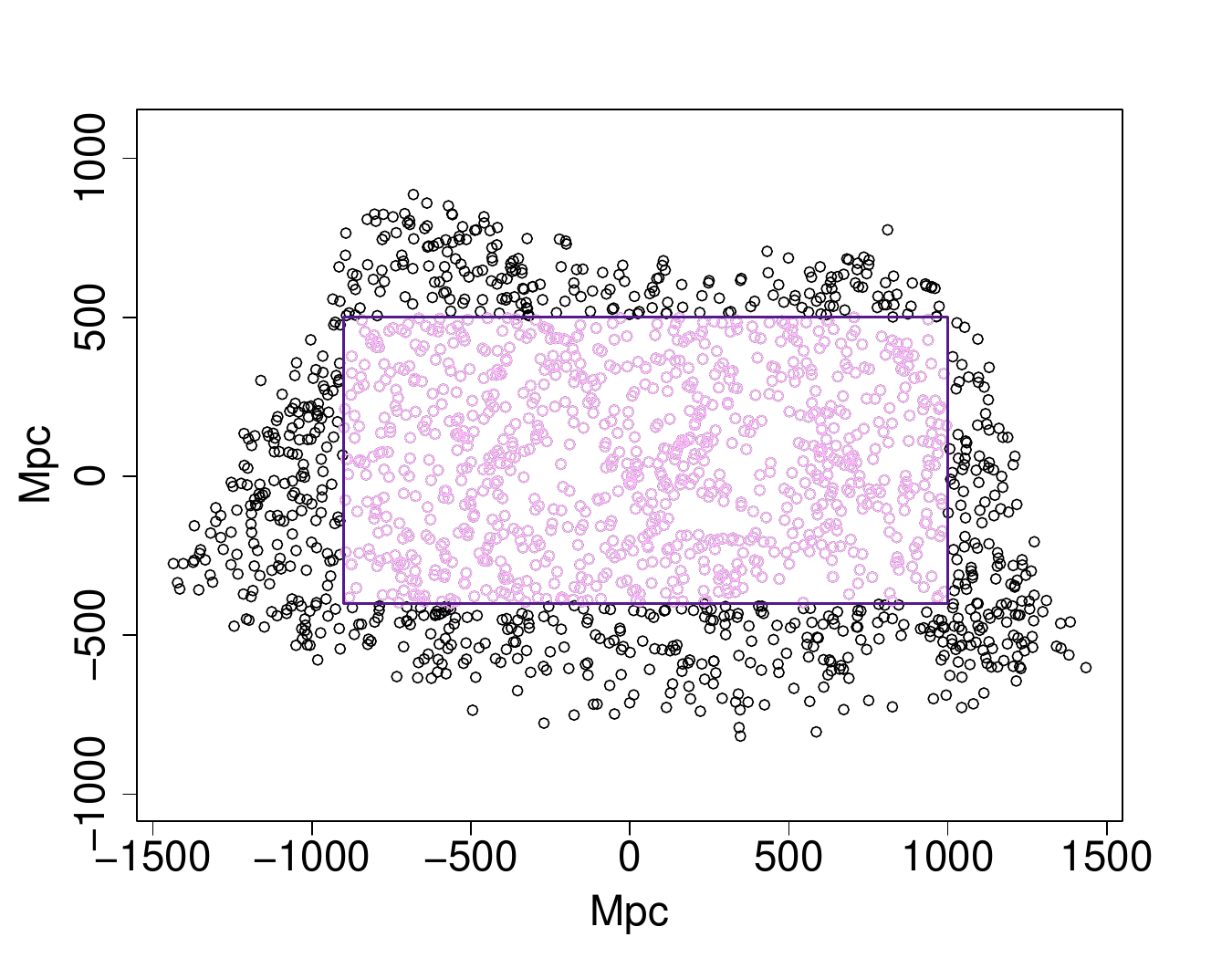}
    \caption{The project-plane distribution of Mg~{\sc II} absorbers in the GR field (standard viewing angle, no tilt) shown as open, black points.  The axes, which are the projected $u_0$ and $v_0$ vectors, are in proper Mpc for the present epoch. The purple, rectangular border defines the control region, and the points lying within the control region are shown as open, pink points. The control borders are defined at: $u_0=(-900$ to $1000)$~Mpc and $v_0=(-400$ to $500)$~Mpc. }  
    \label{fig:ellipse_control_field}
\end{figure}

To determine the significance of the elliptical shells, we generate one million random fields at the control field density, then position the specified elliptical shell at the centre of each random field, and finally compare the elliptical-shell number counts in the random fields with the elliptical-shell number counts in the Mg~{\sc II} field. 
The control field we are using is shown in figure~\ref{fig:ellipse_control_field}.
The borders of the control field are defined: $u_0=(-900$ to $1000)$~Mpc and $v_0=(-400$ to $500)$~Mpc.
When we create the one million random fields, we scale the number of points according to the density in the control field. 
 
\begin{figure} 
    \centering
    \includegraphics[width=1\linewidth]{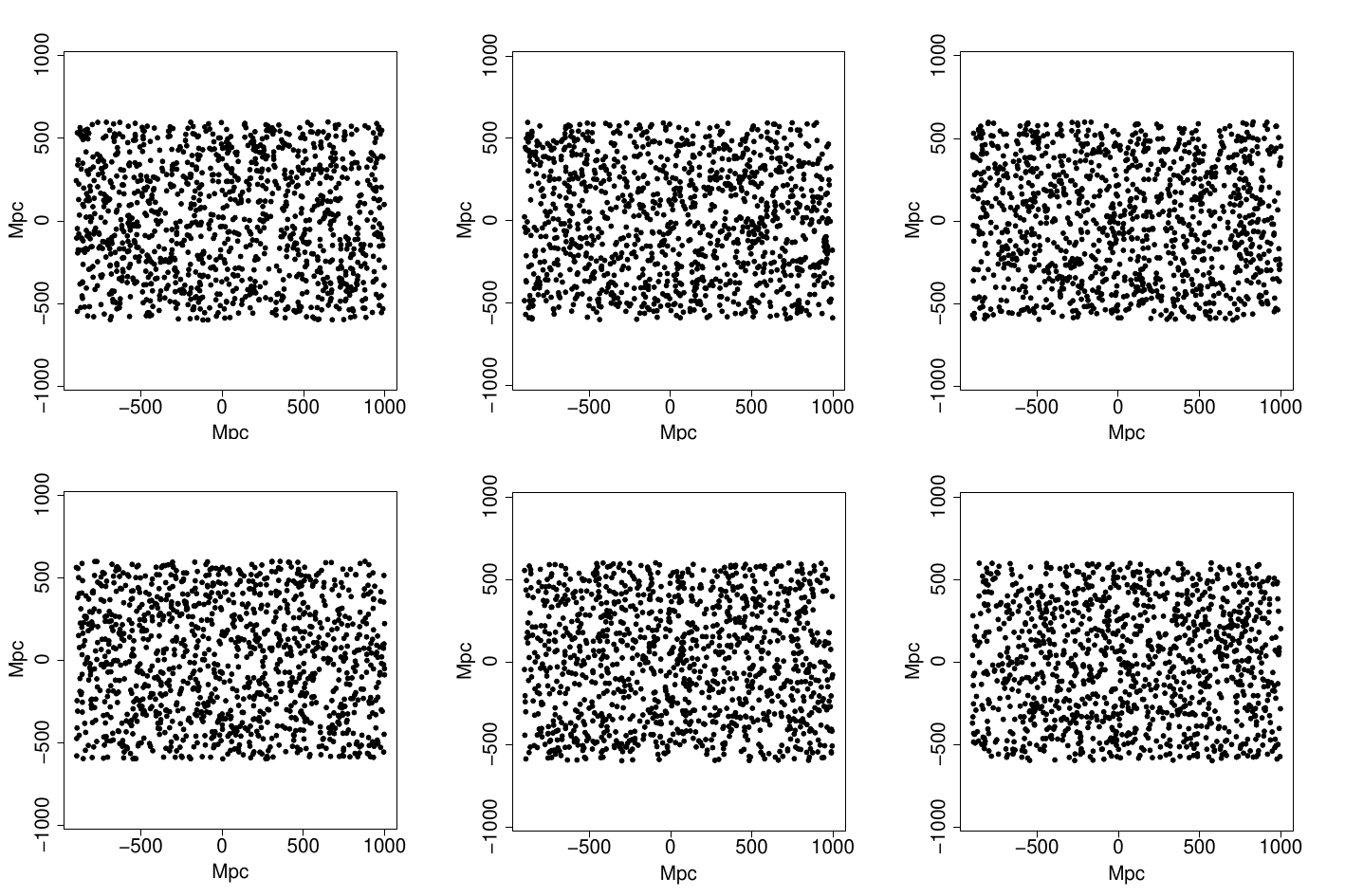}
    \caption{A selection of $6$ random fields that were found to have elliptical-shell number counts equal to or greater than the real GR-predicted elliptical shell in the standard (non-tilted) project-plane FOV; there were $768$ out of one million ($<0.1$ per cent chance) random fields that met this criterion. None of the fields shown here (or others that we looked at) had visually-obvious ellipses within them, indicating that the high number count is likely due to a random alignment of small, dense clusters lying on the elliptical shell, rather than an indication of real ellipse features in the data. The axes in each of these sub-figures are in Mpc, with dimensions equal to the control field in figure~\ref{fig:ellipse_control_field}. }
    \label{fig:6_random}
\end{figure}

In figure~\ref{fig:GR_first_shell} there are $94$ absorber points that fall within the blue ellipse; compared with one million random fields the ellipse has a $3.38 \sigma$ significance, and compared with Poisson expectation the ellipse-count has a $3.28 \sigma$ significance. 
Out of the one million random samples, $768$ had elliptical-shell counts greater than or equal to the observed count.
Notably, however, there were no visually-obvious ellipses upon checking a random sample of these fields (e.g., figure~\ref{fig:6_random}).

The GA+NA ellipse is an a-priori prediction of a GR; in this first simple test we find that the predicted GR is statistically significant.
Let us consider now the following two points: (i) the GR was predicted with very limited data (just the GA and NA) --- any single additional data point could have altered, even a little, the trajectory or position of the GA+NA ellipse-fit; and (ii) the method of assessment with elliptical shells assumes a perfect ellipse with a well-defined thickness --- clearly, the observed data do not conform precisely to these perfect conditions.
Given points (i) and (ii), it now seems impressive that the GA+NA ellipse prediction identifies a significant GR ellipse. 

We can repeat the procedure above but this time in the $5^\circ$ tilted projected plane.
First, the GA and NA need to be projected onto the tilted plane, with a corresponding GA+NA ellipse fit.
The parameters of ellipse in the tilted field are:
\[a = 582; b = 438; \epsilon = 0.247;\theta = 0.304; (x, y) = (11.7, 36.7) \]
where $a$, $b$, $\epsilon$, $\theta$ and ($x, y$) are as defined previously.
The predicted ellipse in the tilted plane is very similar to the predicted ellipse in the standard plane.
In this version, there are $91$ absorber points falling within the elliptical shell; when compared with Poisson expectation the elliptical-shell count has a $3.03 \sigma$ significance, and when compared with one million random fields it  has a $3.12 \sigma$ significance, with $1730$ of those random fields containing ellipse counts greater than or equal to the observed count. 
The GA+NA ellipse is a little less significant here, in the tilted plane, compared with in the standard plane.
Given what we next discuss, this fact further supports the possibility of multiple, overlapping ring features in the data.

\begin{figure} 
    \centering
    \includegraphics[width=1\linewidth]{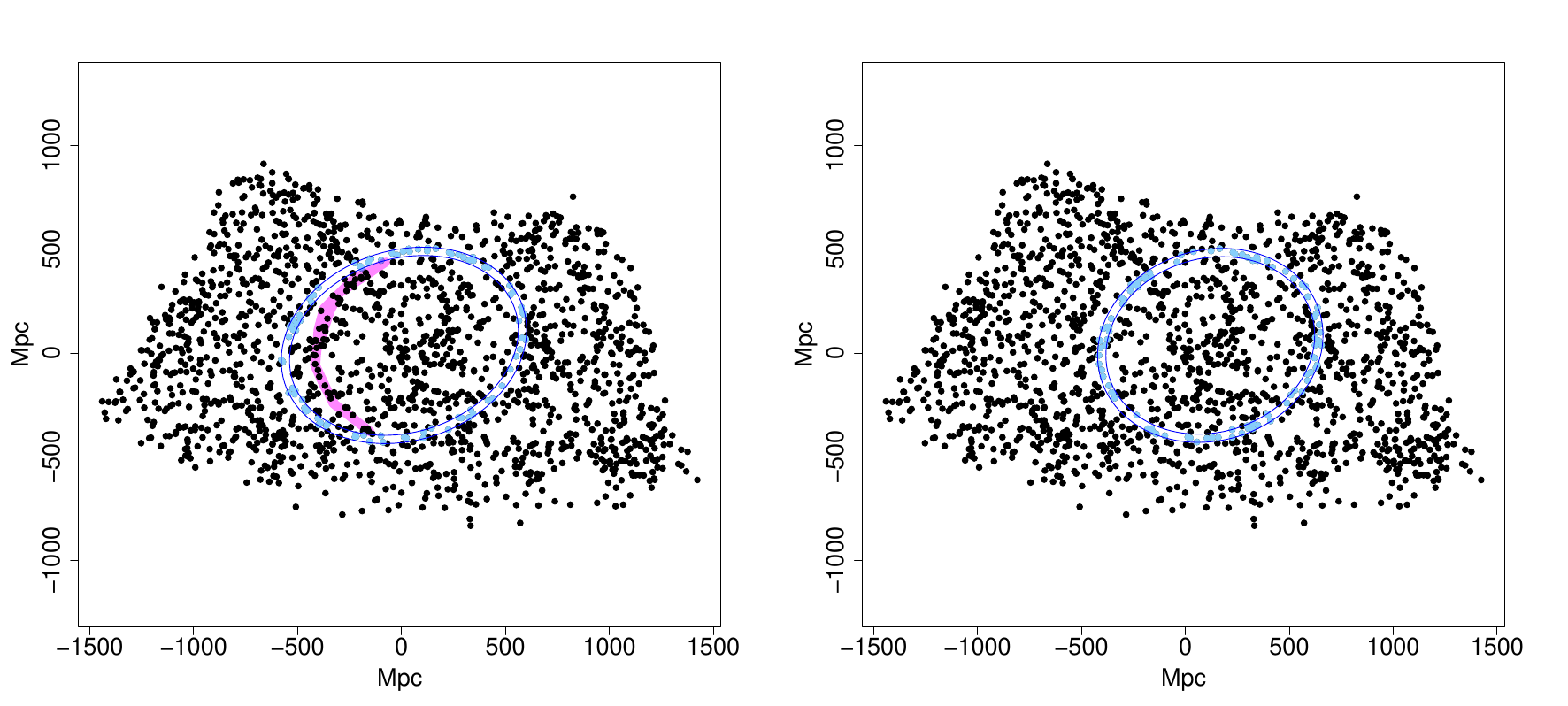}
    \caption{The project-plane distribution of Mg~{\sc II} absorbers in the GR field shown as solid, black points. In this figure, the Mg~{\sc II} absorbers are projected onto the plane perpendicular to $w_0 + 0.1v_0$, which can be thought of as the original line-of-sight projection, modified by a small ($5^\circ$) tilt. The axes, which are the projected $u$ and $v$ vectors, are in proper Mpc for the present epoch, where east is approximately towards the right and north is approximately towards the top, following the convention that $x$ increases towards the right and $y$ increases towards the top.
    Left: the blue elliptical shell corresponds to the GA+NA ellipse in the tilted project-plane field. The solid, blue points are the Mg~{\sc II} absorbers that fall within the elliptical shell. The pink highlighted region indicates the `inner branch' belonging to the visually-identified GR, which also approximately corresponds to filaments identified with FilFinder (see section~\ref{subsec:FilFinder}). 
    Right: the blue elliptical shell approximately corresponds to the visually-identified ellipse in the tilted project-plane field (approximated from adjusting only the $x$ position and semi-major axis of the GA+NA ellipse). The ellipse here is intended to match the ellipse in the left panel of this figure, if using the pink-highlighted inner branch. The solid, blue points are the Mg~{\sc II} points that fall within the elliptical shell. }
    \label{fig:GR_ellipse_predict_and_visual}
\end{figure}

In section~\ref{subsec:FilFinder} we showed the results of the FilFinder algorithm applied to the GR field.
The FilFinder-identified GR filaments appeared consistent with what we determined as the visually-impressive, almost contiguous, roughly circular, GR. 
However, we also noted that the FilFinder algorithm identified branching in the LHS of the GR; if following the trajectory of the outermost LHS branch, then the GR identified here more closely resembles the predicted GR from the GA+NA fit, but if following the trajectory of the innermost LHS branch, then the GR identified here more closely resembles the visually-identified GR.
In the left panel of figure~\ref{fig:GR_ellipse_predict_and_visual} we show the predicted elliptical shell in blue, and highlight the innermost LHS branch corresponding to the visually-identified GR in pink.

We can instead try to estimate this visually-identified GR ellipse by eye.
Clearly, this ellipse must be centred farther right and with a shorter semi-major axis than the GA+NA ellipse (an oversimplification given the range of parameters involved in creating an ellipse, but avoids incorporating too many biased fits).
In this way, the method is heuristic and subject to bias, so in the next section we apply a more objective and robust method of ellipse matching optimisation. 
Nevertheless, for interest, we present what is found with a visually-fit ellipse.

The visually-fit ellipse (for the tilted projected plane) has parameters as follows:
\[a = 530; b = 438; \epsilon=0.214; \theta = 0.304; (x, y) = (120, 36.7), \]
where $a$, $b$, $\epsilon$, $\theta$ and ($x, y$) are as defined previously.
Notice that only the semi-major axis and the $x$ position have changed, for reasons stated above.
By adjusting the ellipse prediction to approximately fit the visually-identified GR, there are now $95$ absorbers that fall into the ellipse yielding in a $4.04 \sigma$ significance compared with Poisson expectations, and $4.16 \sigma$ compared with one million random samples (with only $73$ random fields containing elliptical-shell counts greater than or equal to that observed).
The visually-fitted elliptical shell is shown in the right panel of figure~\ref{fig:GR_ellipse_predict_and_visual}.

In the next section, we apply an ellipse-matching technique which successively tests a range of elliptical shells (positions, orientations, axes) to search for the optimum ellipse signal around the GR.

\subsubsection{Optimum ellipse matching}
\label{subsubsec:optimum_ellipse}

The optimum ellipse-matching method is designed to detect only the optimum ellipse, but there could, of course, be more than one significant ellipse in any one field.
Guided by the ellipse prediction, and also by our visual estimate, we set a wide range of parameters in the ellipse-matching code that allow for the possibility of detecting both current ellipses (the GA+NA and the visually-identified) or any other.
We are applying the optimum ellipse-matching method to the tilted projected-plane Mg~{\sc II} data, which emphasises the visually-identified GR.
The ellipse-matching method here is to optimise the signal of the elliptical shell through systematically checking a range of shells with varying: $x, y$ centres, semi-major and semi-minor axes, and ellipse angles (see table~\ref{tab:ellipse_matching_parameters}).
\begin{table} 
\centering
\begin{tabular}{l|rrrrr}
      & $x$ (Mpc) & $y$ (Mpc) & $a$ (Mpc) & $b$ (Mpc) & $\theta$ ($^\circ$) \\ \hline
Min.      & -10     & -55     & 480     & 420     & -70          \\
Max.      & +140    & +55     & 600     & 520     & +70          \\
Increment & 10      & 10      & 10      & 10      & 10           \\
\end{tabular}
\caption{The ranges of ellipse parameters for optimum ellipse-matching. The ellipse-matching code is designed to test all variations of ellipses with the above parameters, excluding when $b>a$. The code starts with the minimum values (Min.) in each parameter, and then steps through every variation of ellipse with one incremental step (Increment) at a time until the maximum (Max.) of each value is reached. The ellipses defined by the parameters here correspond to the inner ellipses of the elliptical shells (rather than the central ellipses); the outer shells are then defined with the same $x$, $y$, and $\theta$, and with $a+40$~Mpc and $b+40$~Mpc.}
\label{tab:ellipse_matching_parameters}
\end{table}
Note that the ellipses defined by the parameters in table~\ref{tab:ellipse_matching_parameters} correspond to the inner ellipses of the elliptical shells (rather than the central ellipses); the outer shells are then defined with the same $x$, $y$, and $\theta$, and with $a+40$~Mpc and $b+40$~Mpc.

\begin{figure} 
    \centering
    \includegraphics[width=0.8\linewidth]{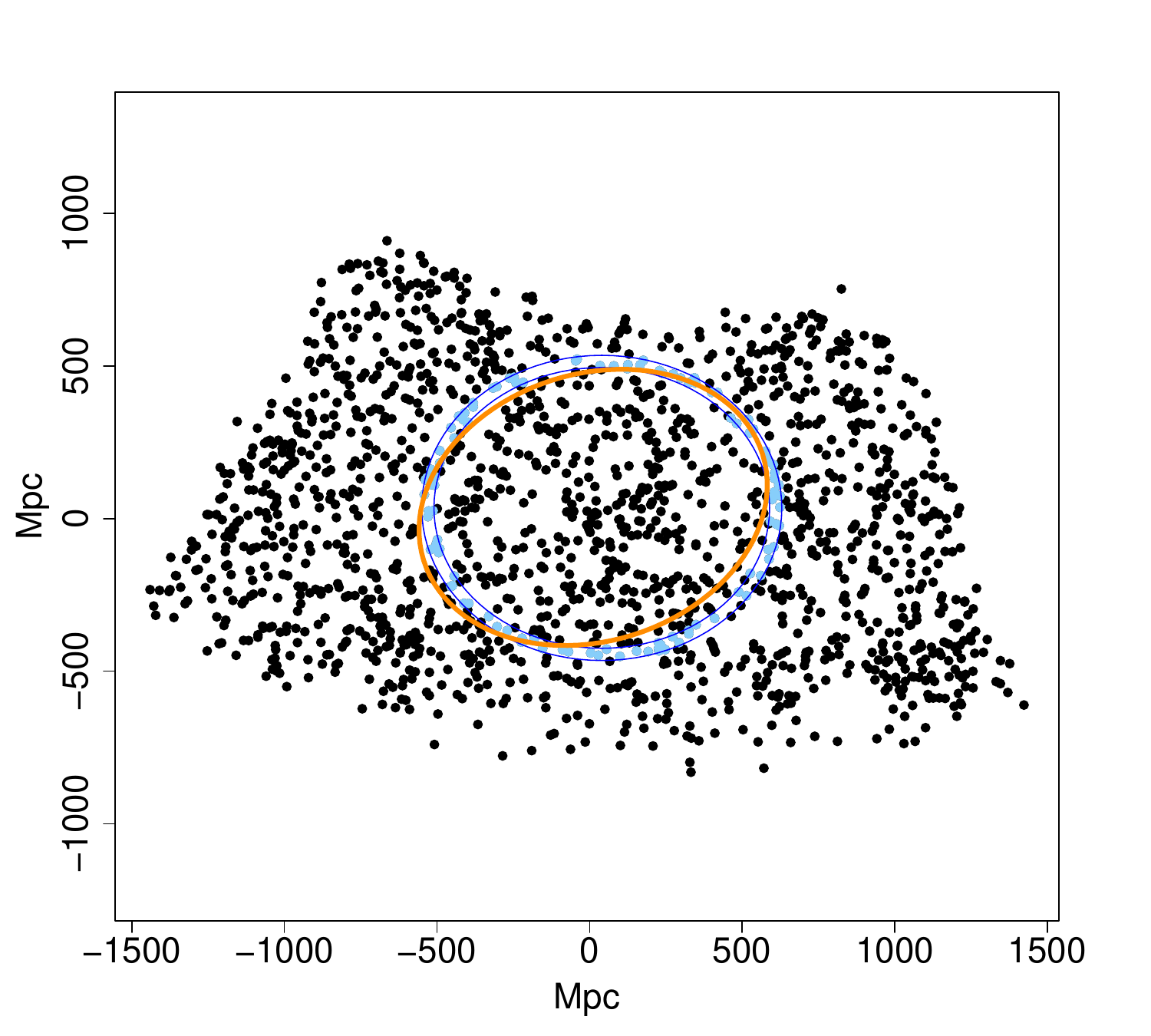}
    \caption{The project-plane distribution of Mg~{\sc II} absorbers in the GR field shown as solid, black points. In this figure, the Mg~{\sc II} absorbers are projected onto the plane perpendicular to $w_0 + 0.1v_0$, which can be thought of as the original line-of-sight projection, modified by a small ($5^\circ$) tilt. The axes, which are the projected $u$ and $v$ vectors, are in proper Mpc for the present epoch, where east is approximately towards the right and north is approximately towards the top, following the convention that $x$ increases towards the right and $y$ increases towards the top.
    The orange ellipse corresponds to the GA+NA ellipse in the tilted project-plane. The blue elliptical shell, and the solid, blue points inside, correspond to the first optimum elliptical shell identified with the ellipse-matching method (with the search parameters specified in table~\ref{tab:ellipse_matching_parameters}). Note that the two ellipses overlap and are quite similar, but the GA+NA ellipse is more elliptical than the optimum elliptical shell. Also, the optimum elliptical shell passes through absorbers below the GA absorber points. Remember, of course, that oversimplifications have been made, such as assuming perfect elliptical shells with a well-defined thickness.}
    \label{fig:GR_ellipse_predict_and_optimum}
\end{figure}

The optimum \emph{central} ellipse identified in the tilted projected-plane field with the ellipse-matching method, using search criteria as specified in table~\ref{tab:ellipse_matching_parameters}, has the following parameters:
\[a = 570; b = 480; \epsilon = 0.158; \theta = 0.0; (x, y) = (40, 35), \]
where $a$, $b$, $\epsilon$, $\theta$ and ($x, y$) are as defined previously. 
Notice here that the ellipticity of this ellipse is rather circular (especially when compared with the other ellipses we have seen up to now).
To calculate the Poisson expectation of the elliptical-shell count, we define a new control field within the project-plane GR field, which has been tilted by $5^\circ$, such that: $u = (-700$ to $+700)$~Mpc and $v = (-500$ to $+500)$~Mpc. 
The GR optimum ellipse here has a Poisson significance of $4.73\sigma$.
By comparing the parameters of this ellipse with the parameters of the original predicted ellipse in the tilted field, we can note their similarities --- see also figure~\ref{fig:GR_ellipse_predict_and_optimum} which shows both ellipses superimposed.
However, while recalling that the GR prediction was based on only the GA and NA members, note that in figure~\ref{fig:GR_ellipse_predict_and_optimum} the optimum ellipse appears to cut through absorbers \emph{above} the GA members. 
We have already mentioned the ambiguity in the southern region of the GR (or the GA region), and we reasoned that multiple overlapping ring features could be the cause of this visual effect.
We bear this in mind as we next attempt to optimise the visually-identified GR ellipse.

As stated above, the optimum ellipse-matching code is designed to detect only the optimum elliptical shell within the search parameters. 
If we wanted to find other, next most-prominent, shells, we would need to forcefully ignore the optimum shell --- doing so is rather complex. 
However, since we suspect there to be a second significant ellipse that coincides with the visually-identified ellipse, and we know that this ellipse fits within the GR ellipse prediction (and optimum), we can reduce the search parameters so that the GR ellipse-prediction shell (and the corresponding optimum shell) cannot be detected (see table~\ref{tab:ellipse_matching_parameters_v2}).
\begin{table} 
\centering
\begin{tabular}{l|rrrrr}
      & $x$ (Mpc) & $y$ (Mpc) & $a$ (Mpc) & $b$ (Mpc) & $\theta$ ($^\circ$) \\ \hline
Min.      &   +60 &  -55   &   480  &   390  & -40         \\
Max.      &  +200 &  +55   &   580  &   490  & +40         \\
Increment &   10 &   10   &    10  &    10  &  10         \\
\end{tabular}
\caption{The range of ellipse parameters for optimum ellipse-matching. The ellipse-matching code is designed to test all variations of ellipses with the above parameters, excluding when $b>a$. The code starts with the minimum values (Min.) in each parameter, and then steps through every variation of ellipse with one incremental step (Increment) at a time until the maximum of each value is reached. In this second (smaller) test, the parameters have been selected in a way which does not allow for the detection of the already-known optimum ellipse (corresponding to the GA+NA ellipse). The ellipses defined by the parameters here correspond to the inner ellipses of the elliptical shell (rather than the central ellipses); the outer shells are then defined with the same $x$, $y$, and $\theta$, and with $a+40$~Mpc and $b+40$~Mpc. }
\label{tab:ellipse_matching_parameters_v2}
\end{table}
 
\begin{figure} 
    \centering
    \includegraphics[width=0.8\linewidth]{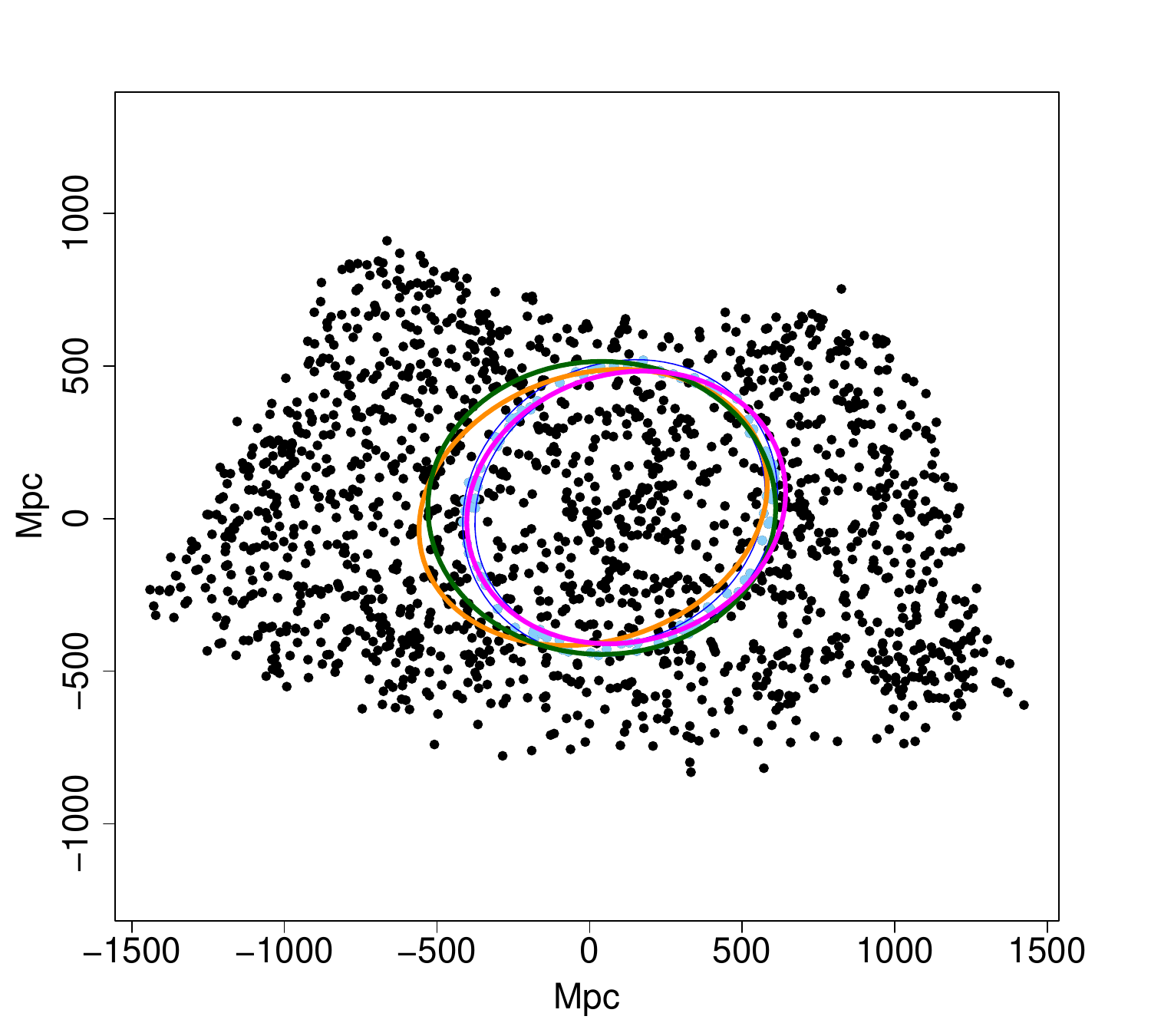}
    \caption{The project-plane distribution of Mg~{\sc II} absorbers in the GR field shown as solid, black points. In this figure, the Mg~{\sc II} absorbers are projected onto the plane perpendicular to $w_0 + 0.1v_0$, which can be thought of as the original line-of-sight projection, modified by a small ($5^\circ$) tilt. The axes, which are the projected $u$ and $v$ vectors, are in proper Mpc for the present epoch, where east is approximately towards the right and north is approximately towards the top, following the convention that $x$ increases towards the right and $y$ increases towards the top.
    The orange ellipse corresponds to the GA+NA ellipse in the tilted project-plane. The green ellipse corresponds to the first optimum ellipse identified with the ellipse-matching method (with the search parameters specified in table~\ref{tab:ellipse_matching_parameters}). The pink ellipse corresponds to the approximately visually-identified GR (based on the GA+NA fit, and adjusting only the $x$ position and semi-major axis length). The blue elliptical shell, and the solid, blue points inside, correspond to the second optimum elliptical shell identified with a second (smaller) set of parameters in the ellipse-matching method (parameters as specified in table~\ref{tab:ellipse_matching_parameters_v2}).  Note here that the approximately visually-identified GR is very close to the second optimum ellipse. The reality of two different versions of the GR, from the GR prediction and from the visual identification, is supported by the detection of two statistically-significant optimum ellipses with the ellipse-matching method.}
    \label{fig:GR_ellipse_visual_fit_and_optimum}
\end{figure}

The optimum \emph{central} ellipse identified with this second (reduced) ellipse matching, using the search criteria as specified in table~\ref{tab:ellipse_matching_parameters_v2}, has the following parameters:
\[a = 510; b = 450; \epsilon = 0.118; \theta = 30; (x, y) = (100, 35), \]
where $a$, $b$, $\epsilon$, $\theta$ and ($x, y$) are as defined previously. 
The GR optimum shell here has a Poisson significance of $4.47\sigma$.
By comparing the parameters of this second optimum ellipse with the parameters of the visually-fit ellipse in the tilted field (which was based on the GA+NA ellipse prediction, then adjusting only the $x$ position and the semi-major axis), we note their similarities --- see also figure~\ref{fig:GR_ellipse_visual_fit_and_optimum} which shows: the second optimum elliptical shell (blue); the visually-fit GR (pink); the GA+NA ellipse (orange); and the GR first optimum shell (green --- all of which are associated with the field tilted 5 degrees).
Earlier we noted that the first optimum ellipse was rather close to the GR ellipse prediction (from the GA+NA), but that the optimum ellipse cut through absorbers \emph{below} the GA members. 
Now note that the second optimum ellipse is very close to the visually-identified GR, but here the optimum ellipse is slightly closer to the GA members than we saw with the first optimum ellipse. 
It seems plausible that multiple overlapping ring features are the cause of the ambiguity in the southern region of the GR (or the GA region) given the overlapping ellipse predictions and optimum ellipses. 

In table~\ref{tab:ellipse_matching_parameters} there are $368, 640$ unique elliptical shells tested; with so many possibilities of detecting significant ellipses, it is now appropriate to consider the `look-elsewhere effect'.
Below, we demonstrate that finding significant ellipses in random data is not difficult to do.
If one wished to refute the importance or significance of an uLSS discovery, one might conclude: \emph{"significant ellipses are easily reproduced in random data, therefore the real structure is simply noise and unimportant".}
Superficially significant ellipses are indeed easily reproduced in random data. 
However, more importantly, in these same random fields we (a) do not find \emph{visually obvious} ellipses, and (b) cannot further confirm the significance of these fields with the use of a second statistical test (for non-random data), namely, the 2D Power Spectrum Analysis (section \ref{subsec:PSA}).
The GR ellipses and the GR field (non-random), in contrast, are statistically significant and, for the GR, visually obvious. 
Additionally, the GR field contains the already statistically-significant, visually-obvious, Big Ring uLSS. 

We have not predicted the statistical likelihood of finding \emph{multiple} significant features with, in particular, ring-like morphologies, in the same redshift slice and FOV; we suspect that the rarity of such an occurrence would be rare indeed, compared with the conservative assessment that we have made here.

By showing that significant ellipses are easily reproduced in random data, while also showing that those same random fields appear entirely random with the powerful 2D PSA method, we demonstrate the complexity involved in carefully assessing LSSs /  uLSSs. 
Note in particularly, the important point that a \emph{single} assessment of a candidate structure is likely to be insufficient for telling the whole story.
By extension, we hope to dispel the misconception that finding these types of patterns in random data is the equivalent of detecting and analysing real cosmological structure.

\begin{figure}
    \centering
    \includegraphics[width=0.6\linewidth]{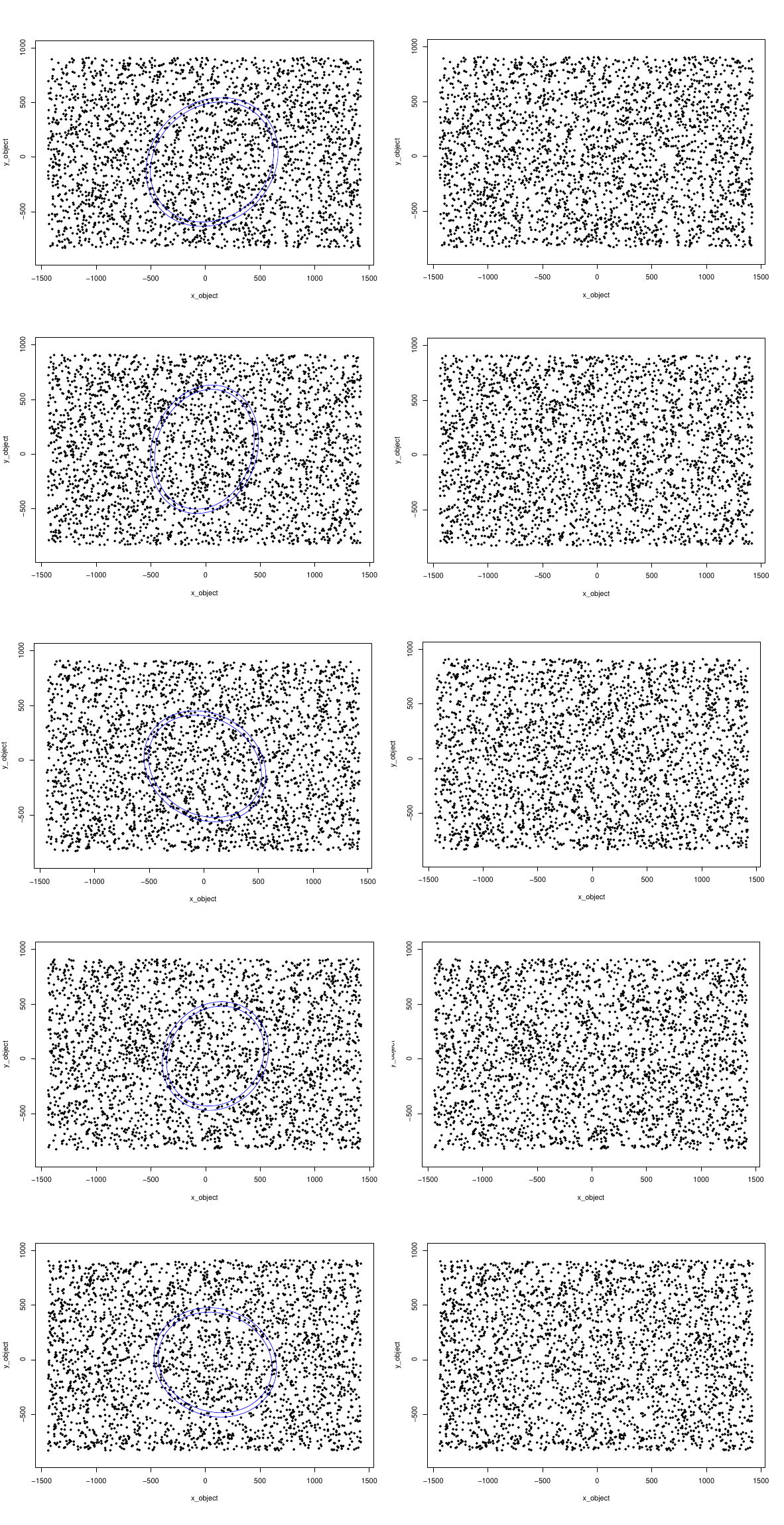}
    \caption{The five random fields (top to bottom) to which we applied the optimum ellipse-matching. The left panels show the random field with the optimum elliptical shell (blue) superimposed, and the right panels show the random field without the added optimum elliptical shell. The side-by-side comparison allows the readers to judge for themselves whether they think any of these elliptical are visually obvious. 
    The optimum shell in the random field in the third row has a Poisson significance $4.92\sigma$, which exceeds the significances of both of the GR optimum shells. The optimum shell in the random field in the fifth row has a Poisson significance $2.89\sigma$, which is the least significant optimum shell from the five random fields. 
    The axes in these sub-figures are in Mpc, with dimensions equal to the control field defined for the tilted project-plane GR field, i.e.,  $u = (-700$ to $+700)$~Mpc and $v = (-500$ to $+500)$~Mpc. }
    \label{fig:5_random_optimum_ellipses}
\end{figure}

For five random fields we perform the optimum ellipse-matching method using the ellipse parameters in table~\ref{tab:ellipse_matching_parameters} (i.e., the larger search area). 
Four out of five of the random fields contain statistically-significant ($\geq 3\sigma$) optimum elliptical shells.
However, only one of the five fields has an optimum shell with a Poisson significance exceeding that of the GR optimum shell.
Figure~\ref{fig:5_random_optimum_ellipses} shows all five random fields with their corresponding optimum elliptical shell highlighted. 
The optimum shell in the random field in the third-row panels of figure~\ref{fig:5_random_optimum_ellipses} has a Poisson significance $4.92\sigma$, which exceeds the significances of both of the GR optimum shells.
The fifth row of figure~\ref{fig:5_random_optimum_ellipses} contains the least-significant optimum shell across the five random fields ($2.89\sigma$).
In all five fields, the optimum shells are not visually obvious, yet, producing these statistically-significance shells with the optimum ellipse-matching method was not difficult to do.

In the next section, we apply the 2D PSA to each of the random fields above.
The 2D PSA results show that all five random fields are completely consistent with random data (no significant clustering detected on any scale).
The combination of both statistical results possibly suggest that the statistically-significant optimum ellipses in the random fields are simply a detection of noise, or otherwise, a result of the `look-elsewhere effect'.

\subsection{Power Spectrum Analysis}
\label{subsec:PSA}
The 2D Power Spectrum Analysis (2D PSA) is a powerful (sensitive) Fourier method for detecting clustering in 2D spatial data on a rectangular plane; see \cite{Webster1976a} and \cite{Webster1976b} for a full description, or section~$5$ in \cite{Clowes1986} for a brief summary. 
We previously used this test to assess the clustering in the GA field:
when applied to a zoomed-in version of the GA field, the 2D PSA results showed significant clustering at wavelengths corresponding to $270$~Mpc;
we interpreted these results as detecting the clustering along the width of the GA.
In this paper, we are now interested in the much larger field containing the GR, so zooming-in is no longer appropriate. 
Additionally, for detecting large-scale clustering, using a larger field will allow for the inclusion of wave-vectors on larger scales.

\begin{figure}
    \centering
    \includegraphics[width=1\linewidth]{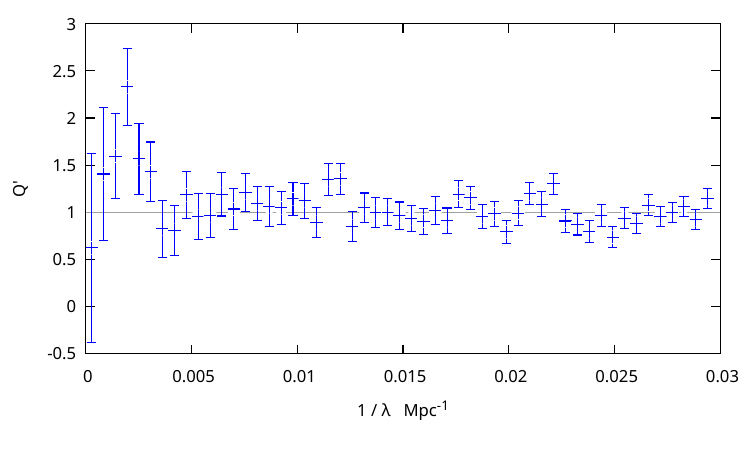}
    \caption{The 2D PSA statistic $Q'$ plotted against $1/\lambda$, with $\lambda$ in Mpc for the project-plane GR field.
The bin size is $5.60 \times 10^{-4}$~Mpc$^{-1}$ and the error bars are $\pm \sigma$. The horizontal line $Q' = 1$ indicates the expectation value in the
case of no clustering. The six left-most points of the plot allow a clustering scale of 
$\lambda_c \sim 320$~Mpc to be identified. The final 
PSA statistic $Q$ for this scale $\lambda_c$ 
corresponds to a detection of clustering at a 
significance of $3.53\sigma$.}
\label{fig:GR_PSA}
\end{figure}

We apply the 2D PSA to the $5^\circ$ tilted project-plane GR field.
The input data is specified with borders: $u=(-900$ to $+900$)~Mpc, and $v=(-450$ to $+550)$~Mpc (i.e., a large area encompassing the GR).
The 2D PSA result is shown in figure~\ref{fig:GR_PSA}.
The combination of the six left-most points of figure~\ref{fig:GR_PSA} corresponds to a $3.53\sigma$ detection for $\lambda_c=320$~Mpc.
When comparing the 2D PSA results to the tilted projected-plane GR field, we might say that this clustering scale corresponds to the areas of `not-voids' --- the areas of
dense filaments that appear to be generally separated by void-like regions (themselves generally spanning a few hundred Mpc).
The large, void-like regions and the dense, arc-like filaments in the GR field creates an impression that the data are `swirly', similarly noted from our FilFinder results earlier in this paper (section~\ref{subsec:initial_checks}).

For comparison, we apply the 2D PSA to random subsets of the FLAMINGO-10K subhalo data.
(The FLAMINGO-10K subhalo data, kindly supplied to us by Till Sawala, is selected at $z=0.7$ for subhaloes in the mass range $1-5 \times 10^{12}$~M$_\odot$.)
First, a volume subset of the FLAMINGO-10K subhalo data is selected to mimic the dimensions of the control field in the tilted project-plane GR field, i.e., a cuboid with dimensions: (x, y, z) = (1800, 1000, 420)~Mpc. 
Then we make a random selection of $964$ subhalo points to match the number of Mg~{\sc II} absorber points in the control field. 
We repeat this to obtain ten samples of $964$ randomly-selected FLAMINGO-10K subhaloes contained within a volume-subset equivalent to the control field within the GR input data. 
Finally, the 2D PSA is applied to each of the ten FLAMINGO-10K fields.
In every case, the FLAMINGO-10K fields were found to be entirely consistent with random expectations, with no significant clustering detected on any scale.

\begin{figure}
    \centering
    \includegraphics[width=1\linewidth]{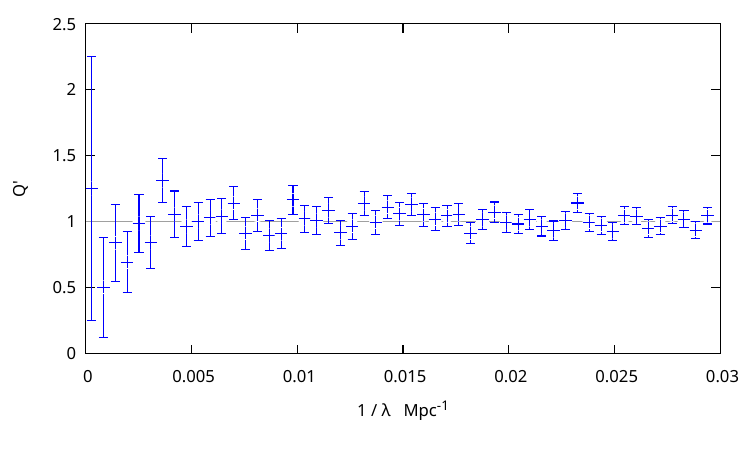}
    \caption{The 2D PSA statistic $Q'$ plotted against $1/\lambda$, with $\lambda$ in Mpc for the random field containing a significant $4.92\sigma$ ellipse. 
The bin size is $5.60 \times 10^{-4}$~Mpc$^{-1}$ and the error bars are $\pm \sigma$. The horizontal line $Q' = 1$ indicates the expectation value in the
case of no clustering. The clustering signal detected in this field are entirely consistent with random expectations.}
    \label{fig:random_PSA}
\end{figure}

Lastly, we apply the 2D PSA to the five random fields in section~\ref{subsubsec:optimum_ellipse}, four of which contained statistically-significant elliptical shells (section~\ref{subsubsec:optimum_ellipse}).
As expected, all five fields were found to be completely consistent with random expectations.
In figure~\ref{fig:random_PSA} we show the PSA result when applied to the random field containing the significant $4.92\sigma$ elliptical shell.

We can now conclude that superficially `significant' elliptical shells detected in the four random fields are random patterns in noise, to be expected as an extreme value, when
there are so many opportunities, or trials (the `look-elsewhere' effect). 
Notably, the reader should consider how the optimum ellipse-matching method, if used incorrectly, might be misinterpreted with a conclusion such as: \emph{"since over-dense ellipses can be found in random data, then over-dense ellipses in real data must also be just patterns in noise"} --- which would of course, be a non sequitur.
However, we have seen that others have made comparable remarks: most prominently, Sawala et al. in their supposed refutation of the \emph{Giant Arc}. 
To remind the reader, the work of Sawala had shown that large, overdense agglomerations of data points (FLAMINGO-10K CDM subhalo points) could be found in simulations and random data, and concluded that this must mean that the Giant Arc is also just a pattern in noise.
(Of course, they did not demonstrate a single `GA-analogue' that simultaneously matched or exceeded the over-density and membership of the real Giant Arc.)
On the contrary, we have now shown that: (i) the two versions of the GR (the GA+NA prediction, and the visually-identified GR) are statistically significant with the ellipse-matching method; (ii) the GR \emph{field} has statistically-significant clustering on large scales (iii) the FilFinder algorithm objectively identifies the filaments belonging to both GR versions; and (iv) direct comparisons with both random data and FLAMINGO-10K cosmological-simulation data were found to be completely consistent with random expectations, even in the case that random, patterns-in-noise were detected.
Combining all of the points above, we have illustrated that a conclusion based on just one simple statistical test is unwise. We have supported further that the GR and the GR field are curious, potentially anomalous, LSS detections, which might in due course prove to have important ramifications for cosmology.

It is worth reiterating here that the 2D PSA is a powerful Fourier method designed to detect clustering in 2D spatial data.
That is, the 2D PSA is \emph{not} designed to detect or identify individual candidate structures.
We have stated previously that finding candidate structures in random data is expected.
However, in these simpler statistical tests, the nuances of the data (and even statistical interpretation) are often missed or improperly handled.
Given that the 2D PSA looks beyond any one individual candidate structure, and assesses the field in its entirety, the results obtained are likely more reliable (compared with other simpler statistical tests, such as FoF, or even ellipse-matching here).
In assessing observational LSS, one should always consider the evidence accumulated from an ensemble of statistical methods, remembering that any one statistical test on its own cannot present the full story. 
  
\section{Summary} 

In this paper we have presented the Giant Ring (GR); a ring-like ultra-large-scale structure (uLSS) at $z\sim0.8$ in the same field that contains the previously-documented Giant Arc (GA) and Big Ring (BR) uLSSs.
The GR appears to be an extension of the GA, and also appears to surround the BR from the perspective along the line-of-sight; the impression is that of two, not-quite concentric, ring-like structures in the sky.
In the same manner as our previous discoveries, the GR was identified with the method of mapping intervening Mg~{\sc II} absorbers in the spectra of background quasars.
However, in contrast to the first two Mg~{\sc II} uLSSs, the GR was \emph{predicted} to exist after noticing an interesting thin, northern filament (the Northern Arc; NA) which looked like it could, with more or enhanced data, connect with the GA to form a giant ring that encompasses the BR.

Guided by the GR prediction, we started to investigate the field.
First, we had begun using a new Mg~{\sc II} catalogue constructed from a pairing between our own Mg~{\sc II} finder applied to the SDSS DR16Q quasars, and the previously-used Anand et al. Mg~{\sc II} catalogue \cite{Anand2021}; we have referred to this as the `high-confidence sample'. 
Next, when considering the predicted on-sky dimensions of the GR, we anticipated that the redshift slice containing the GR would need to be increased.
Additionally, again given the predicted size of the GR, we expected the signal to be lower than what we found with the GA or BR.
Intuitively, the reasoning above can be understood if one considers the filamentary nature of the GR: the large, thin filaments would be easily masked by noise, whereas smaller, denser filaments might be easier to detect. 
We then proceeded to experiment with increasing the redshift slice (only on the near side, so as to not limit the availability of background quasars), and reducing the noise in the field with signal-to-noise (SN) limits in the Mg~{\sc II} lines and \emph{local} continuum estimation.
(A useful aspect of our Mg~{\sc II} catalogue construction is the introduction of \emph{local} continuum estimation, which is a better proxy for the quality of the measured Mg~{\sc II} doublet at that location than the usual \emph{global} continuum estimation.)
Finally, we included the use of the project-plane method which allows the viewing angle of the field to be rotated and tilted. 
Small tilts of the field could enhance the visual impression of real features.
The heuristic processes described above led to the detection of a visually-impressive, almost contiguous, roughly circular, Giant Ring feature (figure~\ref{fig:GR}).
The visually-identified GR is in the redshift slice $0.722 \leq z \leq 0.862$ (equivalent to the usual redshift slice $z=0.802\pm0.060$ with an extension on the near side by $\Delta z = 0.020$), with signal-to-noise (SN) limits of $6, 3$~and~$12$ for the Mg~{\sc II} $\lambda\lambda 2796, 2803$ lines, and the local continuum estimation, respectively, and with a $5^\circ$ tilt of the $u_0, v_0$ vector-plane anti-clockwise with respect to the $w_0$ vector.

We applied a combination of visual-inspection tests and statistical tests to the GR and field containing the GR. 
We summarise the results below.

\subsection*{2D nearest neighbours}
The 2D nearest-neighbours technique was created in analogy with the technique of Voronoi tessellation, in which the smaller Voronoi cells correspond to the higher densities.
The method draws vectors either of actual length or of fixed length between pairs of 2D nearest neighbours, with options to remove points that do not have a nearest neighbour within a specified maximum threshold.
We found that the visual impression of these images enhanced the contrast between clustered areas and voids in the GR field.
The enhanced signal gave a stronger impression of the separation between the GR and the BR (figure~\ref{fig:2d_nn_plots}).

\subsection*{FilFinder}
The FilFinder method is a 2D objective filament-identification algorithm.
We have used this method previously, which identified the filament belonging to the BR, plus an additional `Northern Spur'.
Repeating the procedures set by our previous work, we applied the FilFinder algorithm to the larger field containing the GR (figure~\ref{fig:FilFinder_GR_small_FOV}).
The FilFinder algorithm identified the filaments corresponding to the visually-identified GR.
(Note that here, we are applying FilFinder to a tangent-plane version of the data, so the viewing angle is the standard line-of-sight.)
We saw again the identification of the Northern Spur filament, which extends father than previously, given the larger FOV, and we also found a diagonal Tangential Bar which runs in the north-east direction between the GR and BR. 
The GR identified with the FilFinder algorithm is made up of multiple filaments, some which are thin, `singular', and filamentary, and others which are clustered and `multi-directional'. 
The original NA filament is identified and extends from approximately north-east to north-west. 
We saw branching in the LHS of the GR which we tentatively suggested was potential evidence for multiple overlapping ring features.
The branching creates an outer arc, and an inner arc, which rejoin at the top of the GR (where it is more filamentary) and at the bottom of the GR (where it is more clustered and ambiguous).
We then noted that the outer branch appears to correspond more closely with the GR prediction (otherwise referred to as the GA+NA ellipse), whereas the inner branch appears associated with the visually-identified, almost contiguous, roughly circular, GR.
The two versions of the GR (the prediction and the visually-identified) have not been strictly uncoupled from one another in this paper, but further investigations might reveal two distinct ring features that overlap on the sky, which we shall pursue in future work.

\subsection*{GR analogues}
We demonstrated with a simple visual experiment how one might reproduce `GR analogues' (or, more generally, visually-obvious, ring-like features). 
We generated random data within a cuboid with dimensions equal to the physical size of the GR control field in project-plane coordinates, using the same number of points in the random fields as the number of Mg~{\sc II} absorbers in the control field (i.e., same field density).
We then superimposed artificial spherical clusters in the cuboid of random data, being sure to keep the total number of points the same.
By visually comparing the random data (only) with the random data with added artificial spherical-shell clusters, we demonstrated in a simple manner how one might produce ring-like features in otherwise homogeneous data. 
Perhaps then, this might indicate that the real data is not simply homogeneous on even very large scales, but contains inhomogeneities arising from, possibly, perturbations in the early Universe, or new physics.
We speculate here, but we invite the reader to consider the possibility that such ring-like uLSSs in the Universe are hints towards something beyond the standard cosmological model.

\subsection*{Elliptical-shell assessments}
We assessed the elliptical shells for the parameters the predicted GR and the visually-identified GR.
By defining two concentric ellipses separated by a small value $\Delta r$, we created an elliptical shell with thickness $\Delta r$.
First, we assessed the elliptical shell for the parameters of the predicted GR (from GA+NA).
In the standard project-plane field, the parameters of this ellipse are:
\[a = 588; b = 442; \epsilon=0.248; \theta = 0.278; (x, y) = (5.12, 52.2), \]
where $a$ is the semi-major axis, $b$ is the semi-minor axis, $\epsilon$ is the ellipticity of the ellipse, $\theta$ is the angle of the ellipse anti-clockwise from horizontal in radians, and ($x, y$) is the centre of the ellipse.
There were $94$ absorber points in the elliptical shell with parameters of the GR-predicted ellipse in the standard project-plane FOV; compared with one million random fields this elliptical-shell count had a $3.38 \sigma$ significance, and compared with Poisson expectation the elliptical-shell count had a $3.28 \sigma$ significance. 
Then, we repeated the assessment above, but this time for the GR-predicted ellipse in the $5^\circ$ tilted project-plane FOV.
In the tilted project-plane field, the parameters of this ellipse are: 
\[a = 582; b = 438; \epsilon = 0.247;\theta = 0.304; (x, y) = (11.7, 36.7) \]
where $a$, $b$, $\epsilon$, $\theta$ and ($x, y$) are as defined previously.
Here, there were $91$ absorber points falling within the elliptical shell; when compared with Poisson expectation the elliptical-shell count had a $3.03 \sigma$ significance, and when compared with one million random fields it had a $3.12 \sigma$ significance.
Finally, we estimated an ellipse for the visually-identified GR.
Adjusting only the semi-major axis, and the $x$ centre position of the GR-predicted ellipse, we were able to fit an elliptical shell to what we had approximately visually identified as the GR in the tilted plane, while keeping to a minimum the number of adjusted parameters (to reduce bias).
In the tilted project-plane field, the parameters of the approximately visually-identified GR are:
\[a = 530; b = 438; \epsilon=0.214; \theta = 0.304; (x, y) = (120, 36.7), \]
where $a$, $b$, $\epsilon$, $\theta$ and ($x, y$) are as defined previously.
There were $95$ absorbers that fell into this ellipse, yielding in a $4.04 \sigma$ significance compared with Poisson expectations, and $4.16 \sigma$ compared with one million random samples.

Following on from the elliptical-shell assessment, we apply a method of elliptical-shell matching (or, optimum elliptical-shell matching). 
Continuing with the use of the elliptical shells as above, we then set a wide range of ellipse parameters, such that we systematically progress through a range of shells with varying $x, y$ centres, semi-major and semi-minor axes, and ellipse angles.
The method optimises the elliptical-shell signal through many trials.
The parameters of the first optimum ellipse are: 
\[a = 570; b = 480; \epsilon = 0.158; \theta = 0.0; (x, y) = (40, 35), \]
where $a$, $b$, $\epsilon$, $\theta$ and ($x, y$) are as defined previously.
This first optimum elliptical shell has a Poisson significance of $4.73 \sigma$.
By superimposing the GR-predicted ellipse with this optimum elliptical shell, we note their similarities (figure~\ref{fig:GR_ellipse_predict_and_optimum}).
It is in fact rather remarkable that a statistically-significant GR \emph{prediction} was made from the presence of only a GA and NA which has now been found to be very close to the very significant optimum elliptical shell with the ellipse-matching method. 
Since the optimum elliptical-shell-matching method only finds the first optimum elliptical shell, we made a second, smaller set of ellipse parameters that did not include the first optimum ellipse.
In this way, we were adapting the method to allow for the possibility of optimising what we perceived as the visually-identified ellipse.
The second optimum ellipse identified has parameters:
\[a = 510; b = 450; \epsilon = 0.118; \theta = 30; (x, y) = (100, 35), \]
where $a$, $b$, $\epsilon$, $\theta$ and ($x, y$) are as defined previously. 
This second optimum elliptical shell has a Poisson significance of $4.47 \sigma$.
Similarly, by superimposing the approximately visually-identified ellipse with this optimum elliptical shell, we note their similarities (figure~\ref{fig:GR_ellipse_predict_and_optimum}).
With the optimum ellipse-matching method we provide further support for the reality of two, statistically-significant, possibly distinct, GRs --- one which was predicted from the GA+NA ellipse, and one which was identified visually upon investigating the possibility of a GR.

We next applied the optimum ellipse-matching method to five random fields. 
Given the many trials of ellipses within the range of ellipse-matching parameters, we were easily able to detect statistically-significant elliptical shells in four out of five of the fields.
In one of the random fields, the optimum elliptical shell exceeded the statistical significance of both GR optimums.
However, later, when we applied the powerful 2D Power Spectrum Analysis (2D PSA) to each of the five fields, no significant clustering was detected on any scale (the fields were entirely consistent with random expectations).
Additionally, none of the optimum ellipses identified were \emph{visually obvious} (figure~{\ref{fig:5_random_optimum_ellipses}).

\subsection*{2D power spectrum analysis}
Lastly, we applied the 2D PSA to: the tilted project-plane field containing the GR, ten randomly-sampled FLAMINGO-10K fields, and five random fields (mentioned above).
The GR field was found to have statistically-significant clustering on large scales corresponding to $\sim 320$~Mpc at $3.53 \sigma$. When comparing with the tilted project-plane image (of the GR field), we noted that this $320$~Mpc scale could be determined by
the areas of what are not voids --- the areas of `not-voids'. The image shows large void regions surrounded by dense arc-like clustered areas (not-voids) which may well be characterised by this scale.
In contrast, the ten FLAMINGO-10K fields all appeared to be entirely consistent with random on all clustering scales, and similarly, the five random fields that we discussed above were all also consistent with random expectations. 
The PSA is unlike most other simple statistical tests, such as the FoF or the elliptical-shell-matching method here, in that it assesses the whole field rather than individual candidate structures.
The difficulty in the latter is that, if one is not careful, then patterns in noise can be confused with real structure, or vice versa.

\section{Conclusion}

The Giant Ring was discovered in the same field, and at the same redshift as the two previously-reported uLSSs, the GA and BR. 
Nested rings (such as the BR and GR) seem unlikely to occur in a FLRW homogeneous universe, so perhaps the explanation for these structures lies beyond standard cosmology.
Of recent interest is the alternative model \emph{Conformal Cyclic Cosmology} proposed by Sir R. Penrose \cite{Penrose2014, Meissner2025}.
In this scenario, the collision of supermassive black holes in a previous aeon transfers imprints to the current aeon that might then be perceptible in the CMB \cite{An2018, Gurzadyan2013}. Perhaps these imprints in the early Universe could evolve in such a way that they also become influential in late-time LSS.

In this paper we have presented the preliminary analysis of a Giant Ring on the sky.
The Giant Ring was first \emph{predicted} and then confirmed with the FilFinder algorithm and with elliptical-shell assessments.
In addition, the field containing the GR was found to be significantly clustered (that is, markedly non-random) with the 2D PSA method.
With the FilFinder algorithm and the elliptical-shell assessments, we demonstrated that there appear to be two versions of the GR, both of which are statistically significant.
The two versions of the GR are distinguished by the branching on the LHS of the GR, leading to an `outer' arc and an `inner' arc.
The GR following the outer arc appears closely linked with the GR prediction (from the GA+NA ellipse), and the GR following the inner arc appears closely linked with the visually-identified GR.
The two versions of the GR might be indicating two separate ring features that overlap on the sky.
A visualisation also indicated that, by viewing the southern end of the GR from `underneath' (with the project-plane method), there was a noticeable gap in the redshift distribution, giving the impression of two almost distinct features from this angle. Possibly, in due course, these two features in the redshift distribution will be found to correspond to distinct rings.

In a comparative analysis, random fields containing statistically-significant ellipses were not found to be visually obvious, and their fields were found to be entirely consistent with random expectations with the 2D PSA test. Similarly, all ten of the randomly-sampled FLAMINGO-10K fields were also found to be consistent with random expectations with the 2D PSA test. The superficially `significant' ellipses identified in the random data will be simply patterns in noise, arising from the look-elsewhere effect, given the many trials.

We can now emphasise the importance of carefully assessing LSSs / uLSSs. We have shown that a single candidate structure assessment is insufficient to present the whole story, 
and that one must consider the evidence accumulated from an ensemble of statistical and observational assessments. The proper analysis of LSSs and uLSSs is complex.
We hope now to dispel the misconception that finding these types of patterns in random data is the equivalent of detecting and analysing real cosmological structure. Thus, the accumulated evidence in this paper supports the reality of a GR uLSS.

\acknowledgments
We acknowledge the use of the public R software (v4.1.2)
\footnote{https://www.R-project.org/}.  Our data has depended on the
publicly-available Sloan Digital Sky Survey quasar catalogue and the
corresponding Mg~{\sc II}
catalogues\footnote{https://wwwmpa.mpa-garching.mpg.de/SDSS/MgII/}.
AML was supported by a post-doctoral position from the University of Lancashire.


\begin{thebibliography}{99} 
%
\bibitem{Yadav2010}
J.K. Yadav,  J.S. Bagla, N. Khandai, \emph{Fractal dimension as a measure of the scale of homogeneity}, \emph{Mon. Not. Roy. Astron. Soc.} {\bf 405} (2010) 2009 [arXiv:1001.0617]
%
\bibitem{Lopez2022}
A.M. Lopez, R.G. Clowes and G.M. Williger, \emph{A Giant Arc on the sky}, 
\emph{Mon. Not. Roy. Astron. Soc.} {\bf 516} (2022) 1557 [arXiv:2201.06875]
%
\bibitem{Lopez2024} 
A.M. Lopez, R.G. Clowes, G.M. Williger, \emph{A Big Ring on the sky}, 
\emph{JCAP} {\bf 07} (2024) 055 [arXiv:2402.07591]
%
\bibitem{Sawala2025}
T. Sawala, M. Teeriaho, C. S. Frenk, J. Helly, A. Jenkins, G. Racz, M. Schaller, J. Schaye, \emph{The emperor's new arc: gigaparsec patterns abound in a {\ensuremath{\Lambda}}CDM universe}, \emph{Mon. Not. Roy. Astron. Soc.} {\bf 541} (2025) 1 [arXiv:2502.03515]
%
\bibitem{Gott2005}
J.R. Gott III, M. Jurić, D. Schlegel, F. Hoyle, M. Vogeley, M. Tegmark, N. Bahcall and J.
Brinkmann, \emph{A map of the Universe}, \emph{Astrophys. J.} {\bf 624} (2005) 463 [astro-ph/0310571]
%
\bibitem{Lopez2019}
A.M. Lopez, \emph{Assessing the potential of intervening Mg~{\ sc II} absorbers for cosmology}, \emph{MSc thesis}, \emph{Uni. Cent. Lanc.} (2019)
%
\bibitem{Clowes2012}
R.G. Clowes, L.E. Campusano, M.J. Graham and I.K. S\"ochting, \emph{Two close 
large quasar groups of size $\sim$ 350 Mpc at $z \sim 1.2$}, 
\emph{Mon. Not. Roy. Astron. Soc.} {\bf 419} (2012) 556 [arXiv:1108.6221]
%
\bibitem{Lyke2020}
B. W. Lyke et al., \emph{The Sloan Digital Sky Survey Quasar Catalog: Sixteenth Data Release}, \emph{Astrophys. J. Suppl.} {\bf 250} (2020) 8 [arXiv:2007.09001].
%
\bibitem{Zhu2013}
G. Zhu and B. M\'enard, \emph{The JHU-SDSS metal absorption line catalog: redshift evolution and properties of Mg II absorbers}, \emph{Astrophys. J.} {\bf 770} (2013) 130 [arXiv:1211.6215].
%
\bibitem{Anand2021}
A. Anand, D. Nelson and G. Kauffmann, \emph{Characterizing the abundance, properties, and kinematics of the cool circumgalactic medium of galaxies in absorption with SDSS DR16}, \emph{Mon. Not. Roy. Astron. Soc.} {\bf 504} (2021) 65 [arXiv:2103.15842].
%
\bibitem{Pilipenko2007}
S.V. Pilipenko, \emph{The space distribution of quasars}, \emph{Astron. Rep.} {\bf 51} (2007) 820
%
\bibitem{Koch2015}
E.W. Koch and E.W. Rosolosky, \emph{Filament identification through mathematical morphology},
\emph{Mon. Not. Roy. Astron. Soc.} {\bf 452} (2015) 3435 [arXiv:1507.02289]
%
\bibitem{Zou2021}
H. Zou, J. Gao, X. Xu, J. Ma, Z. Zhou, T. Zhang, J. Nie, J. Wang, S. Zue, \emph{Galaxy clusters
from the DESI legacy imaging surveys. I. Cluster detection}, \emph{Astrophys. J. Suppl. Ser.} {\bf 253}
(2021) 56 [arXiv:2101.12340]
%
\bibitem{Paris2017}
I. P\^aris et al., \emph{The Sloan Digital Sky Survey Quasar Catalog: twelfth data release}, \emph{Astron. Astrophys.} {\bf 597} (2017) A79 [arXiv:1608.06483].
%
\bibitem{Schneider2010}
D. P. Schneider et al., \emph{The Sloan Digital Sky Survey Quasar Catalog. V. Seventh Data Release}, \emph{Astron. J.} {\bf 139} (2010) 2360 [arXiv:1004.1167]
%
\bibitem{Raghunathan2016}
S. Raghunathan, R.G. Clowes, L. E. Campusano, I. K. S\"ochting, M. J. Graham and G. M. Williger, emph{Intervening Mg II absorption systems from the SDSS DR12 quasar spectra}, \emph{Mon. Not. Roy. Astron. Soc.} {\bf 463} (2016) 2640 [arXiv:1608.05112].
%
\bibitem{Lopez2025}
A.M. Lopez, R.G. Clowes, G.M. Williger, \emph{Investigating ultra-large large-scale structures: potential implications for cosmology}, 
\emph{Phil. Trans. Roy. Soc. A} {\bf 383} (2025) 2290 [arXiv:2409.14894]
%
\bibitem{Balazs2015}
L.G. Bal\'azs, Z. Bagoly, J.E. Hakkila, I. Horv\'ath, J. K\'obori, I.I. R\'acz, L.V. T\'oth, \emph{A giant ring-like structure at 0.78 < z < 0.86 displayed by GRBs}, \emph{Mon. Not. Roy. Astron. Soc.} {\bf 452} (2015) 2236 [arXiv:1507.00675]
%
\bibitem{Peebles2022}
P.J.E. Peebles, \emph{Anomalies in physical cosmology}, \emph{Ann. Phys.} {\bf 447} (2022) [arXiv:2208.05018]
%
\bibitem{Binney2024}
J. Binney, R. Mohayaee, J. Peacock, S. Sarkar, \emph{Manifesto: challenging the standard cosmological model},  \emph{Phil. Trans. Roy. Soc. A},  {\bf 383} (2025) 2290
%
\bibitem{topcat}
M.B. Taylor, \emph{TOPCAT \& STIL: Starlink Table/VOTable Processing Software in Astronomical Society of the Pacific Conference Series on Astronomical Data Analysis Software and Systems XIV}, Pasadena, California, USA, October 2004
%
\bibitem{Clowes2013}
R.G. Clowes, K.A. Harris, S. Raghunathan, L.E. Campusano, I.K. Söchting and M.J. Graham,
\emph{A structure in the early Universe at $z\sim1.3$ that exceeds the homogeneity scale of the R-W concordance cosmology}, \emph{Mon. Not. Roy. Astron. Soc.} {\bf 429} (2013) 2910 [arXiv:1211.6256]
%
\bibitem{Park2012}
C. Park, Y-Y. Choi, J. Kim, J.R. Gott III, S.S. Kim, K-S. Kim, \emph{The challenge of the largest
structures in the Universe to cosmology}, \emph{Astrophys. J. Lett.} {\bf 759} (2012) L7 [arXiv:1209.5659]
%
\bibitem{Nadathur2013}
S. Nadathur, \emph{Seeing patterns in noise: gigaparsec ‘structures’ that do not violate
homogeneity}, \emph{Mon. Not. Roy. Astron. Soc.} {\bf 434}, (2013) 398 [arXiv:1306.1700]
%
\bibitem{SylosLabini2026}
F. Sylos Labini, \emph{Hidden role of anisotropies in shaping structure formation in cosmological N-body simulations}, \emph{Phys. Rev. D} {\bf 113} (2026) 2 [arXiv:2508.13765]
%
\bibitem{DiValentino2026}
E. Di Valentino \emph{Cracks in the standard cosmological model: anomalies, tensions, and hints of new physics} (2026) [arXiv:2601.01525]
%
\bibitem{Webster1976a}
A.S. Webster, \emph{The clustering of radio sources — I the theory of power-spectrum analysis}, \emph{Mon. Not. Roy. Astron. Soc.} {\bf 175} (1976a) 61
%
\bibitem{Webster1976b}
A.S. Webster, \emph{The clustering of radio sources — II the 4C, GB and MC1 surveys}, \emph{Mon. Not. Roy. Astron. Soc.} {\bf 175} (1976b) 71
%
\bibitem{Clowes1986}
R.G. Clowes, \emph{Automated quasar detection in the SGP field: a clustering study}, \emph{Mon. Not. Roy. Astron. Soc.} {\bf 218} (1986) 139 
%
\bibitem{Penrose2014}
R. Penrose, \emph{On the gravitization of quantum mechanics 2: Conformal Cyclic Cosmology}, \emph{Found. Phys.} {\bf 44} (2014) 873
%
\bibitem{Meissner2025}
K.A. Meissner, R. Penrose, \emph{The physics of Conformal Cyclic Cosmology} (2025) [arXiv:2503.24263]
%
\bibitem{An2018}
D. An, K.A. Meissner, P. Nurowski, \emph{Ring-type structures in the Planck map of the CMB}, \emph{Mon. Not. Roy. Astron. Soc.} {\bf 473} (2018) 3251
%
\bibitem{Gurzadyan2013}
V.G. Gurzadyan, R. Penrose, \emph{On CCC-predicted concentric low-variance circles in the CMB sky}, \emph{Eur. Phys. J. Plus} {\bf 128} (2013) 2 [arXiv:1302.5162]



\end{thebibliography}
\end{document}